\documentclass[twocolumn,prb,aps,footinbib,amsmath,amssymb,superscriptaddress]{revtex4-2}
\usepackage{graphicx}
\usepackage{dcolumn}
\usepackage{xcolor}
\usepackage{bm}
\usepackage{natbib,hyperref}
\usepackage{amsmath,amssymb,amsfonts,bbold}
\usepackage[normalem]{ulem}
\usepackage{dirtytalk}
\usepackage[caption=false]{subfig}

\def\BiPd{$\beta$-Bi$_2$Pd }
\def\BiPde{$\beta$-Bi$_2$Pd}

\begin{document}
\title{Suppressing Majorana Hybridization in Atomically Assembled Cr Chains on \BiPde}

\author{Jinkyung Kim}
\affiliation{Center for Quantum Nanoscience, Institute for Basic Science (IBS), Seoul 03760, Republic of Korea}
\affiliation{Department of Physics, Ewha Womans University, Seoul 03760, Republic of Korea}
\author{Piotr Kot}
\affiliation{Center for Quantum Nanoscience, Institute for Basic Science (IBS), Seoul 03760, Republic of Korea}
\affiliation{Department of Physics, Ewha Womans University, Seoul 03760, Republic of Korea}
\author{Ryo Ogawa}
\affiliation{Department of Basic Science, University of Tokyo, Meguro, Tokyo 153-8902, Japan}
\author{Atsutaka Maeda}
\affiliation{Department of Basic Science, University of Tokyo, Meguro, Tokyo 153-8902, Japan}
\author{Andreas J. Heinrich}
\email{heinrich.andreas@qns.science}
\affiliation{Center for Quantum Nanoscience, Institute for Basic Science (IBS), Seoul 03760, Republic of Korea}
\affiliation{Department of Physics, Ewha Womans University, Seoul 03760, Republic of Korea}
\author{Nicol{\'a}s Lorente}
\email{nicolas.lorente@ehu.eus}
\affiliation{Centro de F{\'{i}}sica de Materiales CFM/MPC (CSIC-UPV/EHU), 20018 Donostia-San Sebasti\'an, Spain}
\affiliation{Donostia International Physics Center (DIPC), 20018 Donostia-San Sebasti\'an, Spain}
\author{Deung-Jang Choi}
\email{djchoi@dipc.org}
\affiliation{Centro de F{\'{i}}sica de Materiales
        CFM/MPC (CSIC-UPV/EHU),  20018 Donostia-San Sebasti\'an, Spain}
\affiliation{Donostia International Physics Center (DIPC),  20018 Donostia-San
 Sebasti\'an, Spain}
 \affiliation{Ikerbasque, Basque Foundation for Science, 48013 Bilbao, Spain}


\begin{abstract}
Magnetic-atom chains on superconductors can host Majorana zero modes, but in finite chains the two end modes generally hybridize and split away from zero energy. We show that Cr chains assembled atom by atom on the Bi-terminated surface of \BiPd enter a short-localization-length Shiba-chain regime that strongly suppresses this finite-size hybridization. Scanning tunneling spectroscopy reveals a chain-length-driven closing and reopening of the in-gap spectrum, followed by zero-bias modes localized at the chain ends. Green-function calculations reproduce the evolution and show an exchange of electron–hole character across the transition. From the residual splitting we extract a localization length of about three atomic sites, explaining the stability of zero-energy end modes in experimentally accessible chains.
\end{abstract}

\date{\today}

\maketitle

Majorana zero modes (MZMs) are expected to appear at the boundaries of one-dimensional topological superconductors, where their non-local character makes them insensitive to perturbations that do not close the bulk gap~\cite{Yazdani_2023}. Magnetic-atom chains on conventional superconductors provide a particularly direct route to this physics: individual magnetic impurities generate Yu--Shiba--Rusinov (YSR) states inside the superconducting gap, and their hybridization can form Shiba bands that become topological in the presence of magnetic order and spin--orbit coupling~\cite{Beenakker,Yazdani_2013,Yazdani,Ruby_chains,Pawlak_2016,Kim_2018,Li_2018,Choi_2019,Schneider1,schneider_precursors_2022,Beck2023_large_topological_gaps,lo_conte_magnet-superconductor_2024}. However, identifying MZMs in finite chains remains difficult. Zero-bias end states can also arise from trivial Andreev or YSR bound states~\cite{Frolov_2021,PhysRevB.102.245431,PhysRevB.106.104503,kuster_non-majorana_2022,schneider_probing_2023}, so a convincing interpretation requires more than the observation of a localized zero-bias peak.

\begin{figure} [t!]
\centering
\vspace{-0.5cm}
\includegraphics[width=1.0\linewidth]{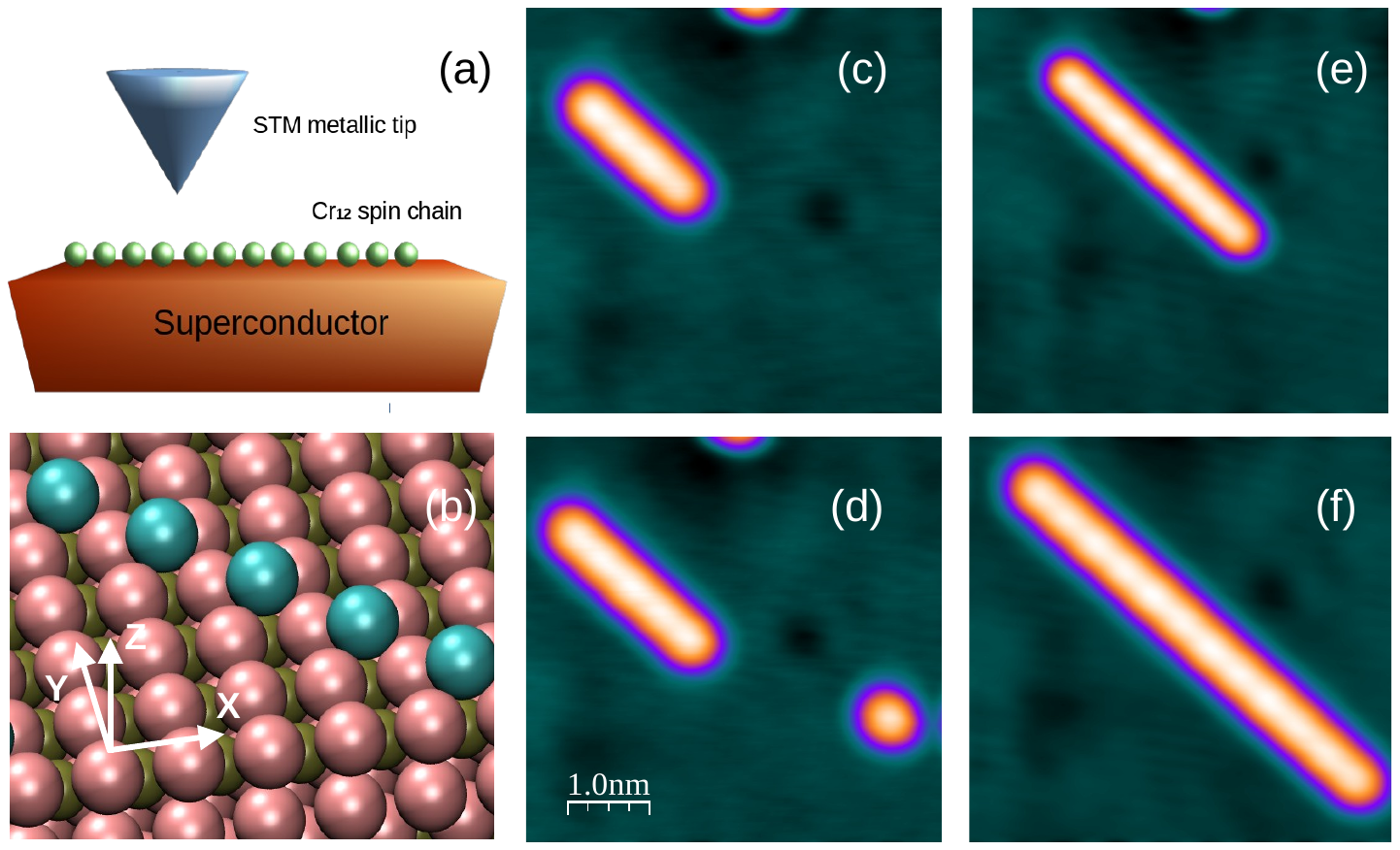}  
\caption{
(a) Schematic of the STM junction used to assemble and probe Cr spin chains on the superconducting \BiPd\ surface. The experiment combines atom-by-atom construction with local tunnelling spectroscopy to follow the finite-size evolution of Shiba-chain end modes. (b) Atomic structure of the chains, with Cr adatoms aligned along the $\langle 110\rangle$ direction of the Bi-terminated surface. The surface consists of a top Bi layer, followed by Pd and Bi layers stacked along the $z$ axis. (c--f) Constant-current STM images of $\mathrm{Cr}_4$, $\mathrm{Cr}_5$, $\mathrm{Cr}_8$, and $\mathrm{Cr}_{12}$ chains assembled atom by atom. STM parameters: $V=3~\mathrm{mV}$, $I=50~\mathrm{pA}$, scan area $5\times5~\mathrm{nm}^2$.
}
\label{figure1}
\end{figure}

A central limitation is the residual hybridization between the two end modes. When the Majorana localization length is comparable to the chain length, the two end-state wave functions overlap and split away from zero energy~\cite{Schneider1,schneider_precursors_2022,Aguado}. This splitting is generally oscillatory because the Majorana wave functions inherit the Fermi-wavelength modulation of the underlying Shiba band~\cite{Choi2022}. Atomically assembled Mn chains on Nb(110) have provided clear evidence for such Majorana precursors: the end states at both ends evolve coherently with chain length, but their energies oscillate strongly and reach zero only for selected chain lengths~\cite{Schneider1,schneider_precursors_2022}. These experiments established the finite-size phenomenology of Shiba-chain Majoranas, but also exposed the need for a material regime in which the localization length is short enough to suppress end-mode hybridization within experimentally accessible chains~\cite{Choi2022,Beck2023_large_topological_gaps,lo_conte_magnet-superconductor_2024}.

The Bi-terminated surface of $\beta$-Bi$_2$Pd offers such a regime~\cite{imai_2012,Choi_2018,Choi2021}. Its low electronic density and strong spin--orbit coupling were predicted to suppress the moir\'e-like oscillatory splitting of Shiba-chain end modes~\cite{Choi2022}, while Cr adatoms on this surface generate YSR states close to the Fermi level that are suitable for forming low-energy Shiba bands~\cite{Choi2021}. Here we assemble Cr chains atom by atom on $\beta$-Bi$_2$Pd and follow their in-gap spectrum with scanning tunnelling spectroscopy as the chain length is increased. We observe a chain-length-driven closing and reopening of the YSR spectrum, followed by zero-bias modes localized at the chain ends. Green-function calculations based on the Bogoliubov--de Gennes formalism reproduce this evolution and show an exchange of electron--hole character across the transition. From the residual splitting of the calculated zero-energy mode we extract a localization length of approximately three atomic sites, demonstrating a short-localization-length Shiba-chain regime in which finite-size Majorana hybridization is strongly reduced.

\begin{figure*}[t!]
\centering
\includegraphics[width=0.7\linewidth]{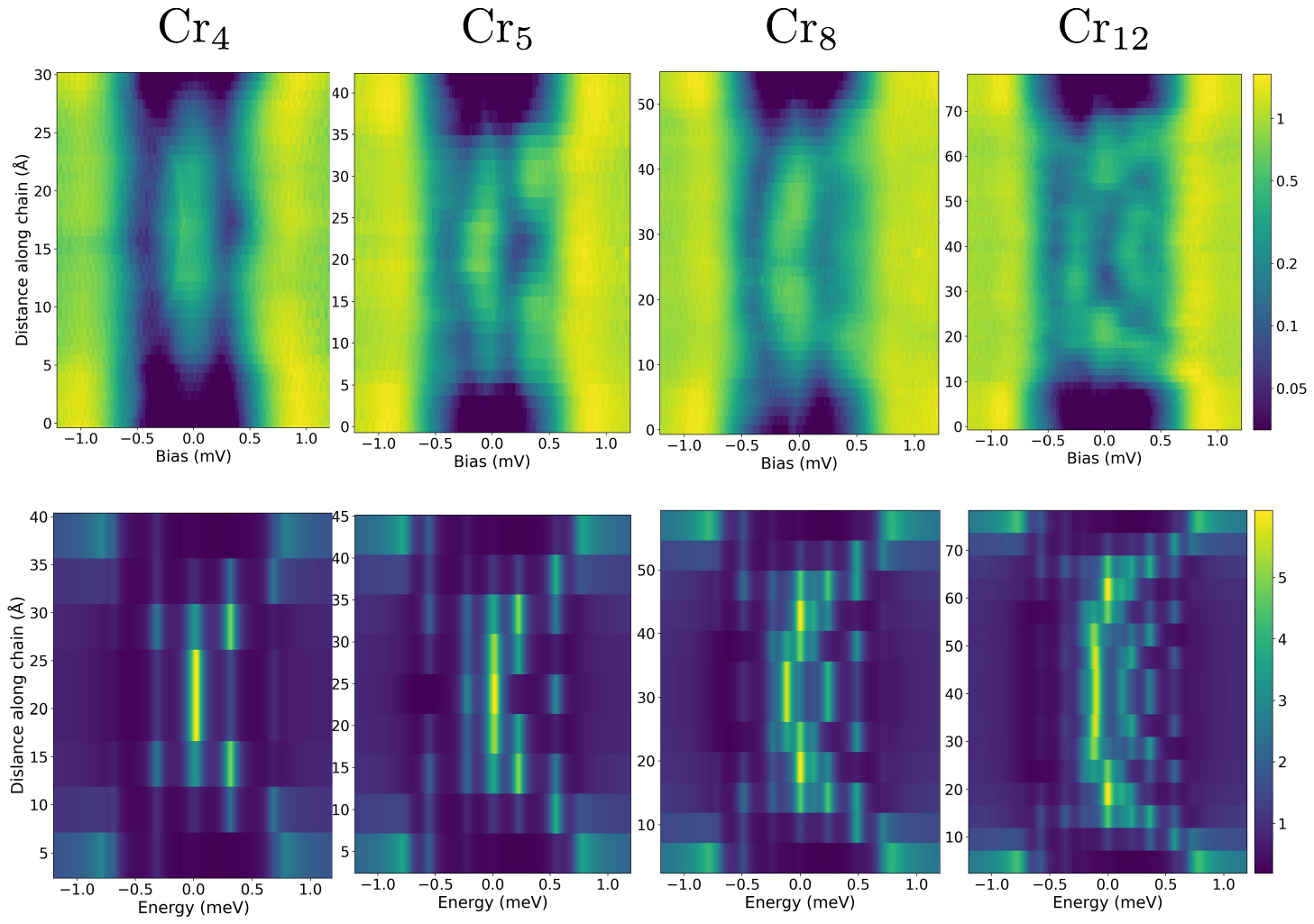}
\vspace{-1.5cm}
\caption{
(a) Spatially resolved differential-conductance maps acquired along Cr$_x$ chains with $x=4,5,8,12$, plotted as a function of position along the chain and applied bias. (b) Computed LDOS showing the same qualitative evolution with increasing chain length as in the experiment in (a): the in-gap YSR states approach zero energy for Cr$_4$, the Shiba-band gap closes near Cr$_5$, and a gap reopens for longer chains. In Cr$_{12}$, the reopened gap is accompanied by zero-bias resonances localized near the two chain ends. This length-controlled closing and reopening of the gap marks the entrance into the confined Shiba-chain regime whose localization length is quantified in Fig.~\ref{figure4}.
}
\label{figure2}
\end{figure*}
Single crystals of \BiPd\ were cleaved \textit{in situ} under ultrahigh vacuum to expose the Bi-terminated superconducting surface. Cr atoms were deposited onto the precooled crystal and manipulated with the tip of a dilution-refrigerator scanning tunnelling microscope operated at $T=30$ mK. Differential-conductance spectra were acquired with a metallic PtIr tip, so that $dI/dV$ directly probes the local density of states of the sample. Additional experimental details are provided in the Supplemental Material.

We assembled Cr chains atom by atom along the $\langle 110\rangle$ direction of the Bi-terminated \BiPd\ surface using lateral STM manipulation [Fig.~\ref{figure1}(a,b)]. The chains were constructed in defect-free regions and their length was increased by adding atoms sequentially to one end of the structure. Figure~\ref{figure1}(c--f) shows representative $\mathrm{Cr}_4$, $\mathrm{Cr}_5$, $\mathrm{Cr}_8$, and $\mathrm{Cr}_{12}$ chains. This controlled length series allows us to follow the evolution from coupled Yu--Shiba--Rusinov (YSR) states to end-localized zero-energy modes in the same material platform.

To interpret the spectra, we calculate the local density of states (LDOS) using a Green-function implementation of Bogoliubov-de Gennes equations for magnetic impurities embedded in an $s$-wave superconducting substrate following Refs.~\cite{Peng_2015,Paaske_2016,Choi2021,Mier2021_b,Jyoti2024}. The clean $\beta$-Bi$_2$Pd surface is described by a BCS Green function including the superconducting gap and Rashba spin--orbit coupling, while each Cr atom is represented by local potential and exchange scattering terms fitted to the single-adatom YSR spectrum~\cite{Choi2021}. The chain Green function is then obtained from Dyson's equation in Nambu space. This approach contains the ingredients relevant for Shiba-band formation and provides a direct theoretical counterpart to the measured $dI/dV$ maps.

Previous work showed that isolated Cr adatoms on \BiPd\ generate YSR states close to the Fermi level and that short Cr chains develop enhanced low-energy spectral weight~\cite{Choi2021}. Calculations further predicted that chains longer than approximately eight atoms should enter a regime in which the in-gap spectrum reopens near zero energy and spatially separated end modes become distinguishable~\cite{Choi2021}. Here we test this prediction directly by measuring spatially resolved differential-conductance maps along chains of increasing length.

Figure~\ref{figure2}(a) shows the measured in-gap spectra for $\mathrm{Cr}_4$, $\mathrm{Cr}_5$, $\mathrm{Cr}_8$, and $\mathrm{Cr}_{12}$. For $\mathrm{Cr}_4$, the spectral weight is concentrated near zero bias, but the experimental resolution does not allow us to resolve a clear mini-gap. Upon adding a fifth atom, the dominant in-gap feature shifts to negative energy, indicating the closing of the low-energy gap. The spectrum evolves further for $\mathrm{Cr}_8$, while for $\mathrm{Cr}_{12}$ a gap centered at zero energy is recovered at the chain center and two zero-bias resonances appear at the chain ends. From the conductance minimum in the central region of the $\mathrm{Cr}_{12}$ chain, we estimate a reopened low-energy gap of $\Delta_{\rm topo}=204\pm50~\mu\mathrm{eV}$. This sizable gap provides the energy scale that confines the end modes~\cite{lo_conte_magnet-superconductor_2024} and is consistent with the short localization length extracted below from the residual splitting~\cite{Peng_2015}.

The calculated LDOS [Fig.~\ref{figure2}(b)] reproduces the same evolution. The YSR resonances approach zero energy for $\mathrm{Cr}_4$, the gap closes near $\mathrm{Cr}_5$, and a reopened gap with end-localized zero-energy states appears for longer chains. The remaining differences between experiment and theory mainly reflect the broader experimental linewidths and the continuous tip trajectory in the measurements, whereas the calculation uses discrete lattice positions and sharper spectral features.
 Experiment and theory therefore show the same length-driven evolution: coupled YSR states approach zero energy, the low-energy gap closes near $\mathrm{Cr}_5$, and a reopened gap with end-localized zero-energy modes appears for longer chains.

\begin{figure}
\includegraphics[width=1.0\linewidth]{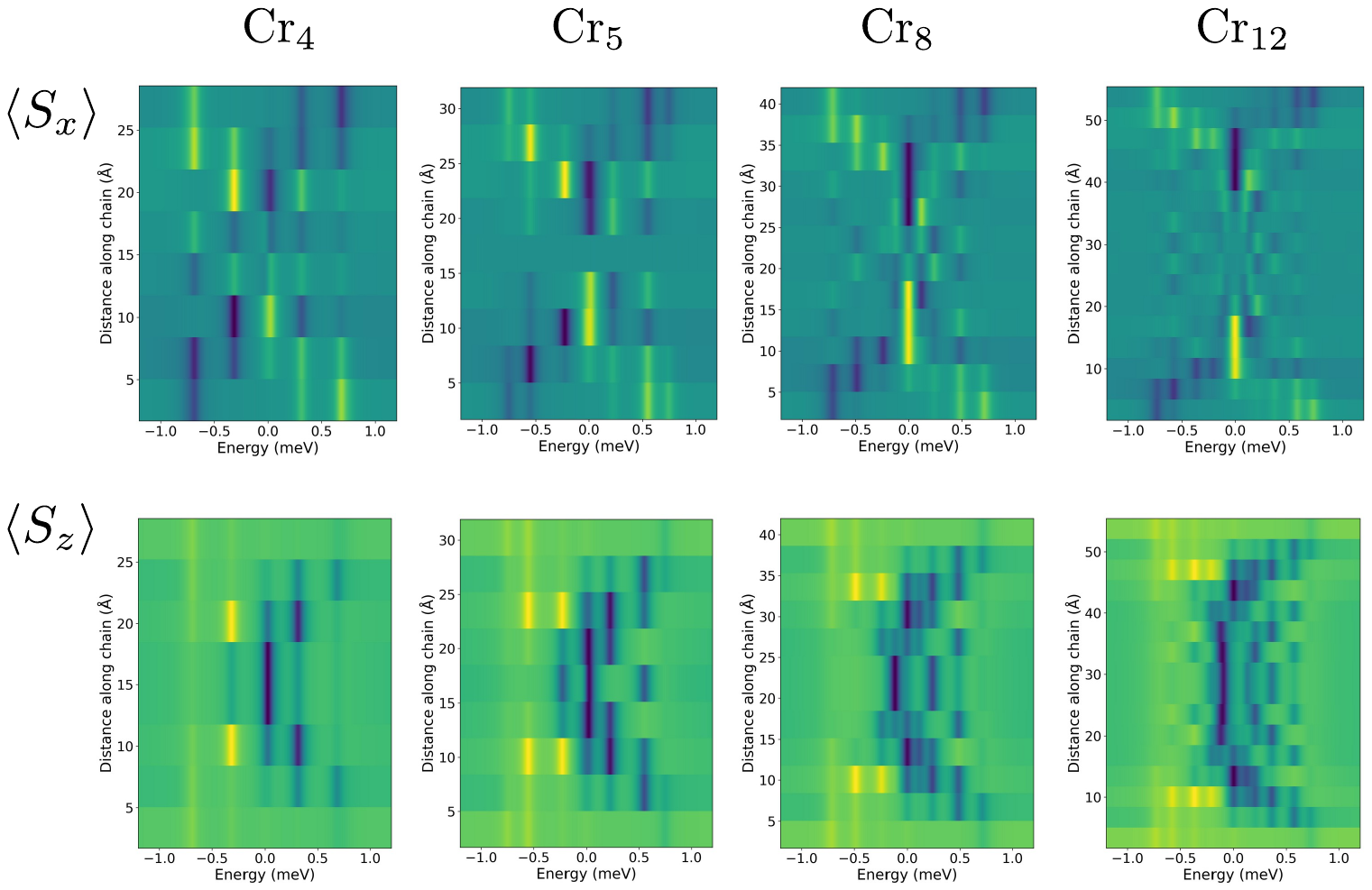}  
\caption{
(a) Calculated in-gap transverse spin component $\langle S_x\rangle$ for $\mathrm{Cr}_x$ chains with $x=4,5,8,12$, plotted as a function of position and energy. Short chains display alternating transverse spin contrast induced by the strong surface spin--orbit coupling. For Cr$_8$ and Cr$_{12}$, the zero-energy spectral weight develops opposite-sign transverse spin features localized near the two chain ends, consistent with the formation of symmetry-related end modes. (b) The longitudinal spin component $\langle S_z\rangle$ show predominantly ferromagnetic alignment, with spin-down and spin-up character associated with electron-like and hole-like branches, respectively. The continuous evolution of $\langle S_z\rangle$ through zero energy reflects the exchange of electron--hole character across the gap closing, while the zero-energy end modes display the electron--hole balance expected for Majorana-like states.
}
\label{figure3}
\end{figure}

Furthermore, the spin texture of the induced Shiba bands gives additional information on the band topology~\cite{Sticlet,Bjornson_2018,Mahdi2020}. Figure~\ref{figure3} shows the spin-resolved LDOS for chains of increasing length. For short chains, the transverse component $\langle S_x\rangle$ displays an alternating pattern induced by the strong surface spin--orbit coupling. For $\mathrm{Cr}_8$ and $\mathrm{Cr}_{12}$, this structure evolves into two opposite-sign features localized near the chain ends, consistent with the formation of spatially separated, symmetry-related zero-energy modes.
The longitudinal component $\langle S_z\rangle$ remains predominantly ferromagnetic across the in-gap spectrum, with spin-down and spin-up character associated with the electron-like and hole-like branches, respectively. As the chain length increases, these branches evolve continuously through zero energy, revealing an exchange of electron--hole character across the gap closing. At zero energy, the calculated end modes display the electron--hole balance expected for Majorana-like quasiparticles, whereas the finite-energy branches remain predominantly electron-like or hole-like.
Together with the gap closing and reopening observed in Fig.~\ref{figure2}, this evolution supports the assignment of the long chains to a topological Shiba-chain regime.

The localization of the in-gap states also presents a marked behavior with chain size. In the shortest chains, the states overlap over the full structure, as seen in Fig.~\ref{figure2}(a). With increasing chain length, the in-gap wave functions reduce their mutual overlap and, consequently, their finite-size hybridization. In the long-chain limit, the two zero-energy modes become independent Majorana-like end states. The relevant question for finite atomically assembled chains is therefore quantitative: whether the localization length is short enough for the two end modes to remain weakly hybridized within experimentally accessible chain lengths.

\begin{figure} [t!]
\centering
\includegraphics[width=1.0\linewidth]{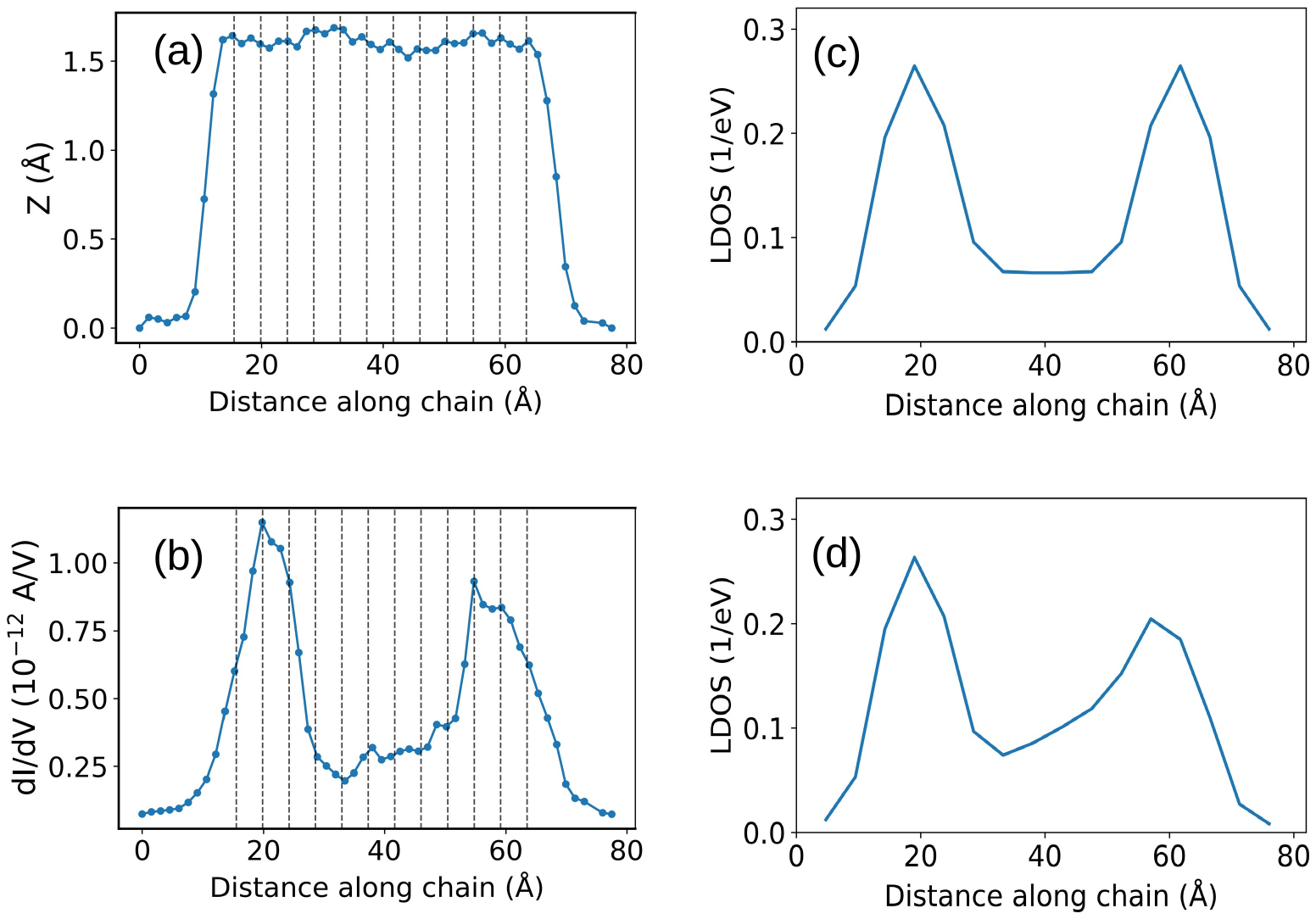}  
\caption{
(a) Constant-current STM topography along the Cr$_{12}$ chain. Vertical dashed lines indicate the Cr atom positions. (b) Zero-bias differential conductance $dI/dV(V=0)$ measured along the same chain, showing spectral weight localized near the two ends and a reduced signal in the central region. (c) Calculated zero-energy LDOS for an ideal chain of identical Cr atoms, yielding nearly mirror-symmetric end profiles. (d) Calculated zero-energy LDOS including a weak linear variation of the local exchange coupling $J$.
}
\label{figure4}
\end{figure}

Figure~\ref{figure4} addresses this point for the $\mathrm{Cr}_{12}$ chain. A constant-current profile over the chain is shown in Fig.~\ref{figure4}(a), while Fig.~\ref{figure4}(b) shows the corresponding zero-bias conductance profile along the same chain. The spectral weight is concentrated near the two ends and is strongly reduced in the central region, indicating compact end-localized modes. The calculation for an ideal chain of identical Cr atoms [Fig.~\ref{figure4}(c)] gives nearly mirror-symmetric zero-energy profiles at the two ends. However, the experimental profile in (b) displays a clear imbalance between the two edge amplitudes.

This imbalance is confined to the end regions. The central part of the chain remains comparatively weakly affected, suggesting that the asymmetry originates from a local or slowly varying microscopic perturbation that modifies the spatial distribution of the end-mode spectral weight without removing the zero-energy state~\cite{Moore}. To test this possibility, we introduce a variation of 6\% between one end atom and the central atom in their respective Kondo exchange interactions with the substrate, keeping the other half of the chain unperturbed. The resulting calculated LDOS [Fig.~\ref{figure4}(d)] reproduces the observed distortion of the end-state profile.  Thus, a small variation of the local coupling to the superconducting substrate is sufficient to account for the measured asymmetry while preserving the zero-energy character of the end state~\cite{Moore,schneider_probing_2023}.

Key to the presence of well-defined zero-energy edge modes is the localization induced by the minigap~\cite{lo_conte_magnet-superconductor_2024}. Numerically, we estimate the confinement by extracting the residual hybridization of the two end modes in the $\mathrm{Cr}_{12}$ chain. In the calculation, the artificial Dynes broadening can be reduced to $\Gamma_D=5~\mu\mathrm{eV}$, revealing that the nominally zero-energy mode consists of two peaks separated by
$\delta\epsilon=16.8~\mu\mathrm{eV}$.
The corresponding localization length can be estimated from the standard expression for hybridized Majorana end states~\cite{Aguado},
\begin{equation}
\delta\epsilon =
2\,\frac{\hbar^{2}}{m} k_{F}\,
\frac{e^{-2L/\xi_M}}{\xi_M}
\cos(k_F L),
\end{equation}
where $L=12\times4.75=57$~\AA\ is the geometric chain length and
$k_F\simeq0.18~a_0^{-1}$ for Bi$_2$Pd~\cite{Herrera}. Since no moir\'e pattern is observed for Cr chains on Bi$_2$Pd~\cite{Choi2022}, we take
$\cos(k_F L)\simeq1$. Using the calculated splitting yields
$\xi_M\simeq10.5$~\AA, corresponding to an edge-state localization of three lattice sites, in very good agreement with the edge-mode extension shown in Fig.~\ref{figure4}(b).

The short localization length is consistent with the sizable reopened gap at the chain center. In a one-dimensional topological superconductor the end-mode localization length scales as
$\xi_M\sim\hbar v_{\rm eff}/\Delta_{\rm topo}$, where $\Delta_{\rm topo}$ is the reopened topological gap and $v_{\rm eff}$ is the velocity of the low-energy band from which the Majorana mode derives. In Shiba chains, however, $v_{\rm eff}$ is strongly renormalized by the hybridization of the magnetic-impurity states with the superconducting substrate, allowing Majorana end modes to localize on a length scale much shorter than the coherence length of the host superconductor~\cite{Peng_2015}. Thus, the measured low-energy gap of approximately $204\pm50~\mu{\rm eV}$ can produce a localization length of only a few atomic sites when combined with the narrow, strongly renormalized Shiba band obtained in the Green-function calculation. The independently extracted value $\xi_M\simeq10.5$~\AA\ is therefore consistent with the velocity-renormalization mechanism~\cite{Peng_2015} expected for magnetic-atom Shiba chains.


The large $\Delta_{\rm topo}$ gap and consequently the short zero-energy mode localization length places Cr/$\beta$-Bi$_2$Pd in a regime distinct from Nb-based Shiba chains~\cite{Schneider2,schneider_precursors_2022}, where finite-size hybridization produces pronounced length-dependent splitting of the end modes in good agreement with previous calculations~\cite{Choi2022}. 

Microwave-assisted tunnelling provides an additional probe of the reopened gap and of the compact zero-energy spectral weight. We applied a 44~GHz modulation to the tip bias and measured the differential conductance as a function of both $V_{\mathrm{RF}}$ and $V_{\mathrm{DC}}$ at different positions along the $\mathrm{Cr}*{12}$ chain (Supplemental Material, Fig.~S5). At the chain end, the zero-bias mode remains centered at zero bias for small $V*{\mathrm{RF}}$ and gradually merges with the background as the modulation amplitude is increased. At the chain center, where the zero-bias conductance is suppressed in the absence of RF driving, increasing $V_{\mathrm{RF}}$ enhances the low-energy spectral weight. This behavior is consistent with microwave-induced broadening of the zero-energy end modes and of the bands defining the reopened gap: the edge signal is reduced, while spectral weight from the gap edges fills the conductance minimum at the chain center. The RF response therefore supports the picture of compact zero-energy modes confined by a sizable reopened gap.

In summary, these measurements and calculations establish Cr/$\beta$-Bi$_2$Pd as a short-localization-length Shiba-chain platform. The evidence does not rely on a single zero-bias peak, but on the combined observation of a chain-length-driven gap closing and reopening, end-localized zero-energy modes, exchange of electron--hole character across the transition, compact real-space profiles, and a residual splitting corresponding to $\xi_M\simeq 10.5$~\AA. This localization length, of only about three atomic sites, is the key physical difference with respect to Shiba-chain systems dominated by length-dependent Majorana splitting. It allows the two end modes to remain weakly hybridized in chains that can be assembled and probed atom by atom. Our results therefore identify a material route to suppressing finite-size Majorana hybridization, providing an atomic-scale testbed for controlled studies of localized topological end modes and their residual coupling.

\pagebreak

\clearpage
\newpage


\onecolumngrid

\setcounter{page}{1}
\setcounter{equation}{0}
\setcounter{figure}{0}
\renewcommand{\theequation}{S.\arabic{equation}}
\renewcommand{\thefigure}{S\arabic{figure}}
\renewcommand*{\thepage}{S\arabic{page}}
\renewcommand{\thetable}{S\arabic{table}}

\renewcommand{\thesection}{S\arabic{section}}
\renewcommand{\thesubsection}{\thesection.\arabic{subsection}}
\renewcommand{\thesubsubsection}{\thesubsection.\arabic{subsubsection}}
\noindent
\textbf{\large Supplementary Information for:}\\

\textbf{\large Suppressing Majorana Hybridization in Atomically Assembled Cr Chains on \BiPde}
\\

Jinkyung Kim, Piotr Kot, Ryo Ogawa, Atustaka Maeda, Andreas J. Heinrich$^\mathrm{a}$, Nicol\'as Lorente$^\mathrm{a}$, Deung-Jang Choi$^\mathrm{a}$\\
\\
$^\mathrm{a}${Corresponding authors:\\ heinrich.andreas@qns.science, nicolas.lorente@ehu.eus and djchoi@dipc.org}

\section{Experimental details}

Single crystals of $\beta$-Bi$_2$Pd were grown following the procedure described in Ref. \cite{imai_2012}. The selected crystal exhibited a superconducting transition temperature 
$T_c=5.2$ K. A clean Bi-terminated surface was obtained by in-situ cleavage under ultrahigh vacuum. Chromium atoms were deposited onto the precooled surface at $T\sim 20$ K, ensuring isolated adatoms.

All measurements were performed in a home-built dilution-refrigerator STM operated at 
$T=30$ mK at the IBS Center for Quantum Nanoscience. A metallic PtIr tip was used so that the differential conductance 
$dI/dV$ directly reflects the local density of states of the substrate. Conductance spectra were acquired with a lock-in amplifier using a 30 $\mu$V excitation at 330 Hz. 

Lateral manipulation of individual Cr atoms was carried out by approaching the STM tip to one side of the adatom until a junction resistance of a few tens of k$\Omega$ (typically achieved at 3 mV bias and tens of nA) was reached. With the feedback loop open, the tip was then translated laterally to controllably drag the atom to the target position.
\section{Summary of the theoretical methods}

We model the Cr spin chains using the dilute-impurity framework of Ref.~\cite{Choi2021}. 
This approximation is justified by density-functional-theory (DFT) calculations showing that 
no Cr $d$-states cross the Fermi level, thereby suppressing charge-transfer processes~\cite{Choi_2018}. 
The superconducting substrate is described through Nambu Green's functions for an $s$-wave BCS superconductor 
following Refs.~\cite{Flatte_1997,Flatte_2000}. A Rashba spin--orbit interaction is added in the local basis to 
account for the strong surface SOC of Bi-terminated $\beta$-Bi$_2$Pd.

The BCS Green's function is parametrized using a Fermi velocity $v_F = 0.15$ (Hartree atomic units) and a Dynes 
broadening $\Gamma = 0.05$\,meV. These parameters yield density-of-states (DOS) peaks narrower than in experiment, 
but facilitate tracking the evolution of in-gap Yu--Shiba--Rusinov (YSR) states with increasing chain length. 
The substrate is treated as a single-gap superconductor with $\Delta = 0.76$\,meV~\cite{imai_2012,Herrera}, and a 
normal-state DOS $N = 0.037/\mathrm{eV}$, chosen to reflect the five valence electrons of Bi.

Magnetic impurities are incorporated through the local Hamiltonian
\begin{equation}
    \hat{H}_{\mathrm{imp}}
    = \sum_j \left(U_j + J_j\,\vec{S}_j \cdot \vec{s}_j\right),
    \label{Kondo}
\end{equation}
with a single compact orbital per site and strictly local exchange and potential scattering~\cite{Pientka2013}. 
The classical spin of impurity $j$ is 
$\vec{S}_j$  and the electron spin operator is strictly local at site $j$ representing the usual Kondo Hamiltonian.
Parameter values are taken from fits to single-Cr YSR spectra: 
$U_j = 5.5$\,eV (capturing the electron--hole asymmetry) and $J_j = 2.5$\,eV. For more details please
refer to Ref.~\cite{Choi2021}.

The strong Rashba coupling $\alpha_R \approx 1.8$\,eV$\cdot$\AA\ is obtained from DFT, Ref.~\cite{Choi2021} and consistent with 
known Bi-terminated surfaces.

We calculate the
Green's function using Dyson’s equation
where $\hat{H}_{imp}$  in the Nambu basis becomes the self-energy and the free-electron BCS Green's function~\cite{Flatte_1997,Mier2021_b} is the bare one. The LDOS is obtained from the electron block of the retarded Green's function,
$\rho(\mathbf{r},E)=-(1/\pi)\operatorname{Im}\operatorname{Tr} G^R_{ee}(\mathbf{r},\mathbf{r};E)$,
and the spin-resolved LDOS is obtained by inserting the corresponding spin
Pauli matrices in the trace.
\section{In-gap spectroscopy of the {Cr$_{12}$} chain}

\begin{figure*} [h!]
\centering
\includegraphics[width=.70\linewidth]{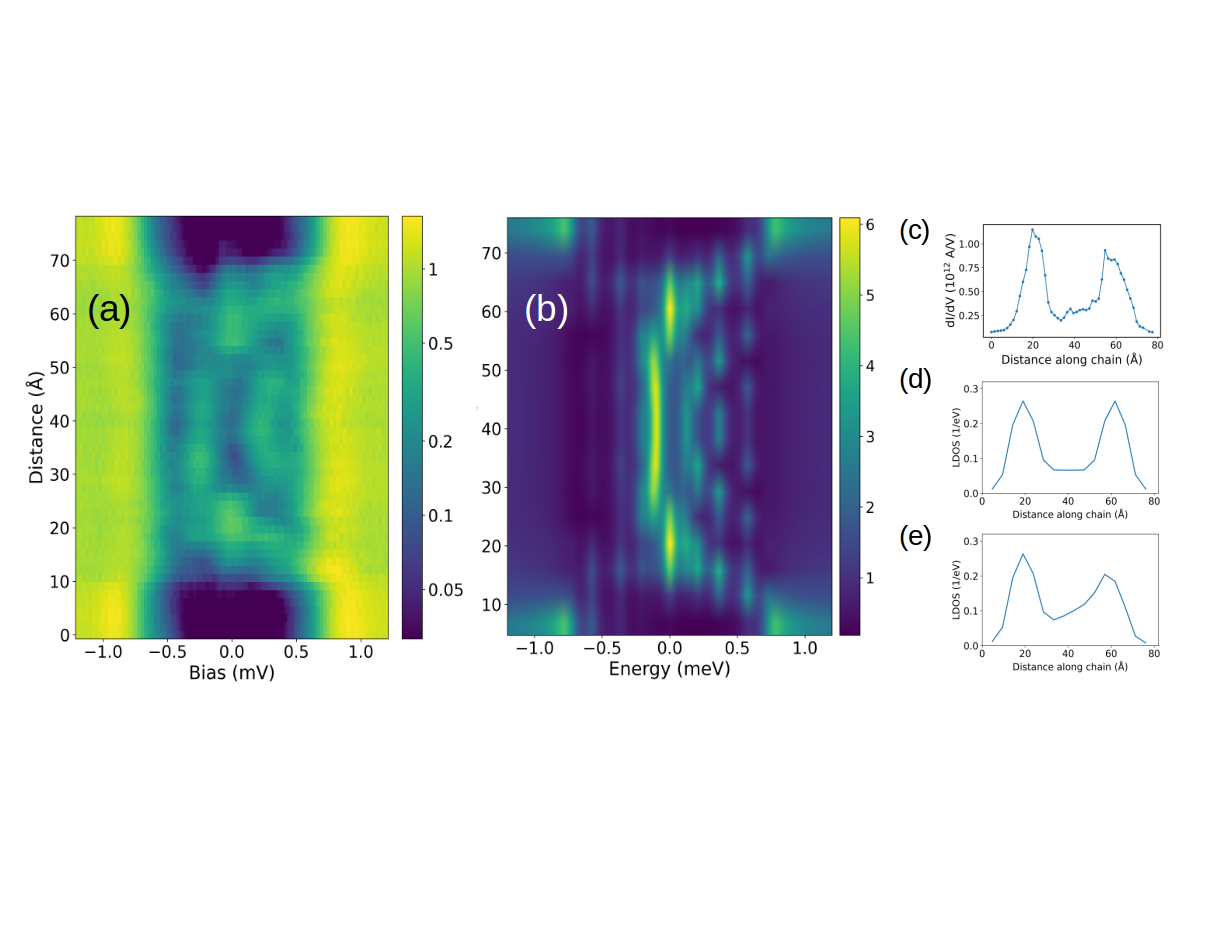}  
\vspace{-1.5 cm}\caption{(a) $dI/dV$ map as a function of the applied bias and the position of the STM over the Cr$_{12}$  chain. (b) Calculated LDOS using interpolation to smooth the discrete character of the calculation. (c) Profile along zero-energy of the experimental data (a). (d) Profile along the calculated (b). (e) Equivalent profile with a linear accumulative perturbation of 1\% along the spin-chain half.}
\label{figureS4}
\end{figure*}

Figure~\ref{figureS4} shows the spatial distribution of the differential conductance as a function of applied bais (a) and the LDOS as a function of quasiparticle energy (b) over the Cr$_{12}$ chain. The profile along the $V=0$ line (c) clearly shows a pronounced asymmetry between edges as discussed in the main text, that does not appear in the calculation (d) unless a variation of the local couplings $J_i$ between each Cr total spin, $\vec{S}_i$ and the electronic spin density at the atomic site $i$ of the superconductor, $\vec{s}_i$, takes place. 

Figure~\ref{figureS4} (e) qualitatively resembles the experimental profile, Fig.~\ref{figureS4} (c), and for that we used $J_i=2.5$ eV for the first 6 atoms  and then decreases linearly by 1\% per atom over the second half of the chain, reaching $J_{12}=2.35$ eV, i.e. a total variation of 6\%. The idea of applying the variation was to show that a small linear perturbation such as one produce by strain on the spin chain could already explain the experimental data.

\section{Spin texture of the in-gap states}

The spin texture of the in-gap states is shown in Fig.~3 of the main text. The zero-energy spectral weight displays a staggered transverse spin texture consistent with the spin structure expected for Majorana-like end modes, Ref.~\cite{Sticlet}. 

\begin{figure*} [h!]
\centering
\includegraphics[width=0.7\linewidth]{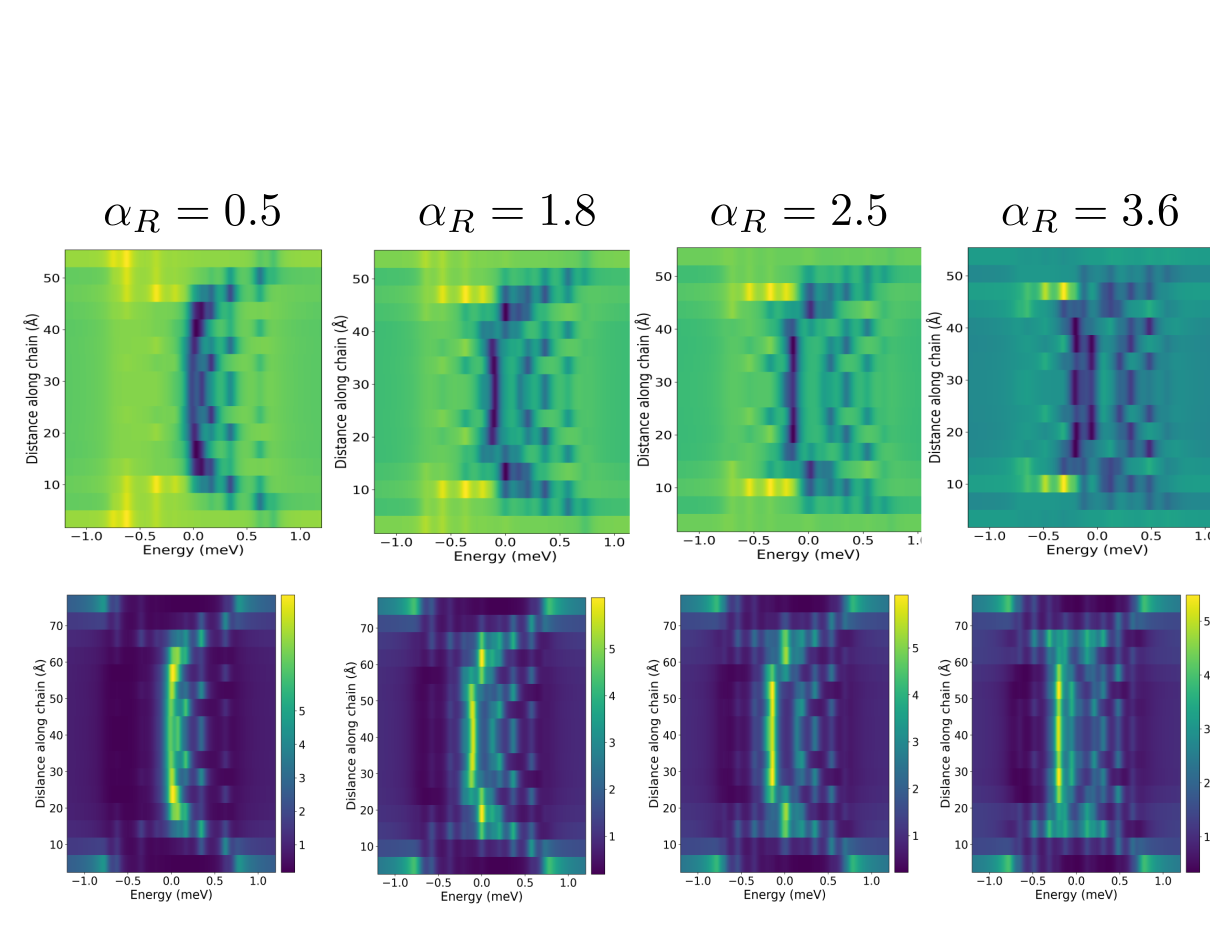}  
\caption{ The upper line shows the behavior of $\langle S_z\rangle$ as Rashba coupling $\alpha_R$ is increased and the lower line corresponds to the LDOS. the The behavior of the in-gap states and particular of the topological gap is not monotonous with the Rashba coupling $\alpha_R$. We find a maximum of the topological gap with $\alpha_R$ close to the Rashba couplings estimated for the Bi$_2$Pd surface.}
\label{figureS2}
\end{figure*}

The spin texture provides an additional consistency check of the gap inversion. This is seen in the crossing of zero energy by spin-down states. This same behavior is more clearly seen in Fig.~\ref{figureS2} where the increase of the Rashba coupling $\alpha_R$ leads to a stabilization of the topological phase and thus to a larger topological gap. 
In experiments such as Ref.~\cite{Yazdani2017}  and \cite{Wiesendanger2019}, the change from opposite to same spin-contrast at ±E is taken as evidence for a topological gap reopening.

Figure~\ref{figureS2} shows the effect of the Rashba coupling $\alpha_R$ on the in-gap states of the Cr$_{12}$ chain. The upper shows the behavior of $\langle S_z\rangle$ as Rashba coupling $\alpha_R$ is increased, while the lower line is the LDOS. We can see that the Rashba coupling is instrumental in closing an reopening the gap, leading to a topological quantum phase transition. The critical value of the Rashba coupling is difficult to determine from these calculations. In our previous work~\cite{Mier2021_b}, we found that very small values of $\alpha_R$ were enough to reopen the gap, producing a non-trivial topological invariant in the correct regions of the phase space. We also noticed that the geometry of the actual system is of great importance to capture the correct in-gap electronic structure.

In the present work, we have increased the coupling to a region were the gap closes again, $\alpha_R=3.6$~eV-\AA. 

At low coupling, the spin-down structure is confined to positive-valued energies. As the coupling is increased, the gap closes and spin-down states cross to negative energies, signaling the reopening of the gap in good agreement with the LDOS evolution. 

\section{Majorana localization and chain length}

Reducing the Dynes parameter in the calculation reduces the artificial broadening
of the in-gap states. This allows us to resolve the residual splitting of the
nominally zero-energy end modes and to estimate their localization length.
The RF-induced redistribution discussed in Fig.~S5 is qualitatively reproduced
by increasing the Dynes broadening, supporting the interpretation of the RF
drive as an additional dynamical broadening mechanism.

Another important test can be done on the theory data by reducing the Dynes parameter. When this broadening attaints $\Gamma_D=$ 5$\mu$eV we find that the zero-energy mode is actually split and the distance between the positive and negative peaks is $\delta \epsilon=16.8\; \mu$eV. This shows that albeit small there is interaction between the localized edge states of the 12-atom chain.  From this splitting we can estimate the exponential confinement of the zero energy modes~\cite{Aguado}: 
\begin{equation}
    \delta \epsilon= 2 \frac{\hbar^2}{m} k_F \frac{e^{-2 L/\xi_M}}{\xi_M} \cos{k_F L},
\end{equation}
where the chain length is  $L=12\times 4.75=57$~\AA\ and $k_F\approx0.18$~a$_0^{-1}$ for Bi$_2$Pd~\cite{Herrera}. Since we do not see any moir\'e effect for Cr on Bi$_2$Pd~\cite{Choi2022}, we assume that $\cos{k_F L}\approx 1$ and we obtain $\xi_M \approx 10.5$~\AA. This is consistent with detecting the zero-energy mode for relatively small chains.

\section{Length behavior of the spin chains: strain}

The mismatch between the equilibrium inter Cr distance and the $\sqrt{2}a$ distance imposed by the substrate leads to substantial accumulation of strain that can give rise to stacking fault and rupture of the spin chains as their length increases. The Cr-Cr distance of a free dimer Cr$_2$ is $d=1.58$~\AA\ noticeably shorter than the $\sqrt{2}a=4.75$~\AA\ of the Cr-Cr distance in the chains. When adsorbed, the Cr atom acquires charge and the balance of forces with the surface leads to a different $d$ distance, but we expect this to be closer to 2~\AA\ rather than the $4.75$~\AA\ of the chain. We conclude that there is strain that will limit the achievable length of the chain.

Furthermore, the accumulation of strain can be inhomogeneous in the chain leading to small geometrical distortion that might explain the difference in Cr-substrate interactions that can explain the difference in signal between the two edges, Fig. 3. In this case, the interaction will be local and will keep the topology of the perfect chain.

\section{Topological gap determination}

\begin{figure} [h!]
\centering
\includegraphics[width=0.3\linewidth]{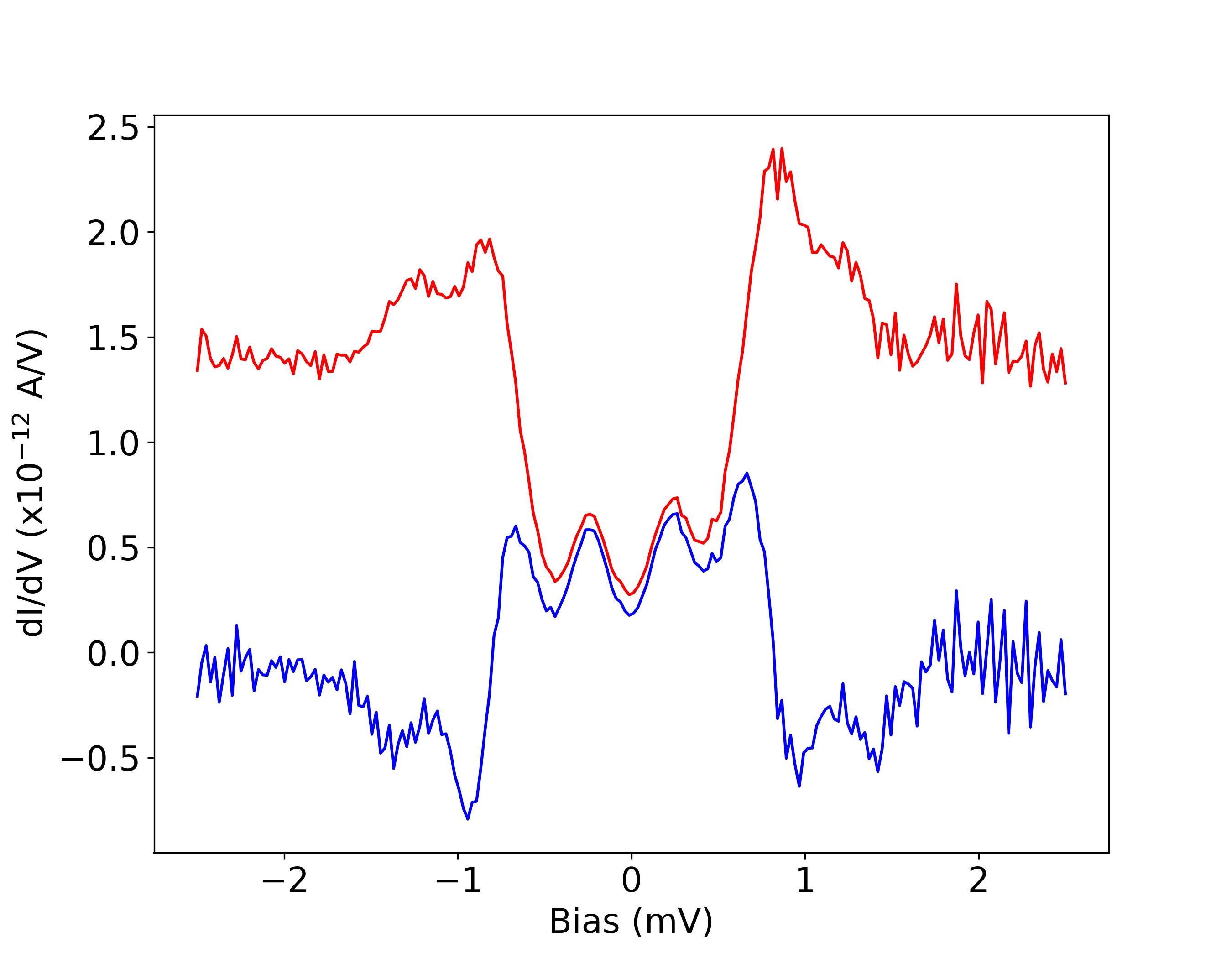}  
\caption{The red curve is the full conductance clearly showing the topological gap. The blue curve is the red curve after subtraction of a $dI/dV$ curve taken in the clean base superconductor. The subtraction emphasizes the in-gap structure without perturbing it.}
\label{figureS3}
\end{figure}

From the experimental data we can estimate the topological gap $\Delta_{\mathrm{topo}}$ as the smallest gap about zero energy between in-gap states. To do this we subtract the $dI/dV$ of the pristine surface from the middle of the chain $dI/dV$ as shown in Fig.~\ref{figureS3}. The pristine surface is reproducibly obtained at the clean superconductor at both edges of the spin chain leading to a very good zero signal if we subtract them, showing the quality of our data.

From Fig.~\ref{figureS3} we obtain that $\Delta_{\mathrm{topo}}=204\pm50$~$\mu$eV. This gap is substantially larger than the 84 $\mu$eV obtained for $\alpha_R=1.8$~eV-\AA\ and 146 $\mu$eV obtained for $\alpha_R=2.5$~eV-\AA\ from Fig.~\ref{figureS2}.

\section{Radio-frequency excitation of the in-gap states}

\begin{figure*} [h!]
\centering
\includegraphics[width=1.0\linewidth]{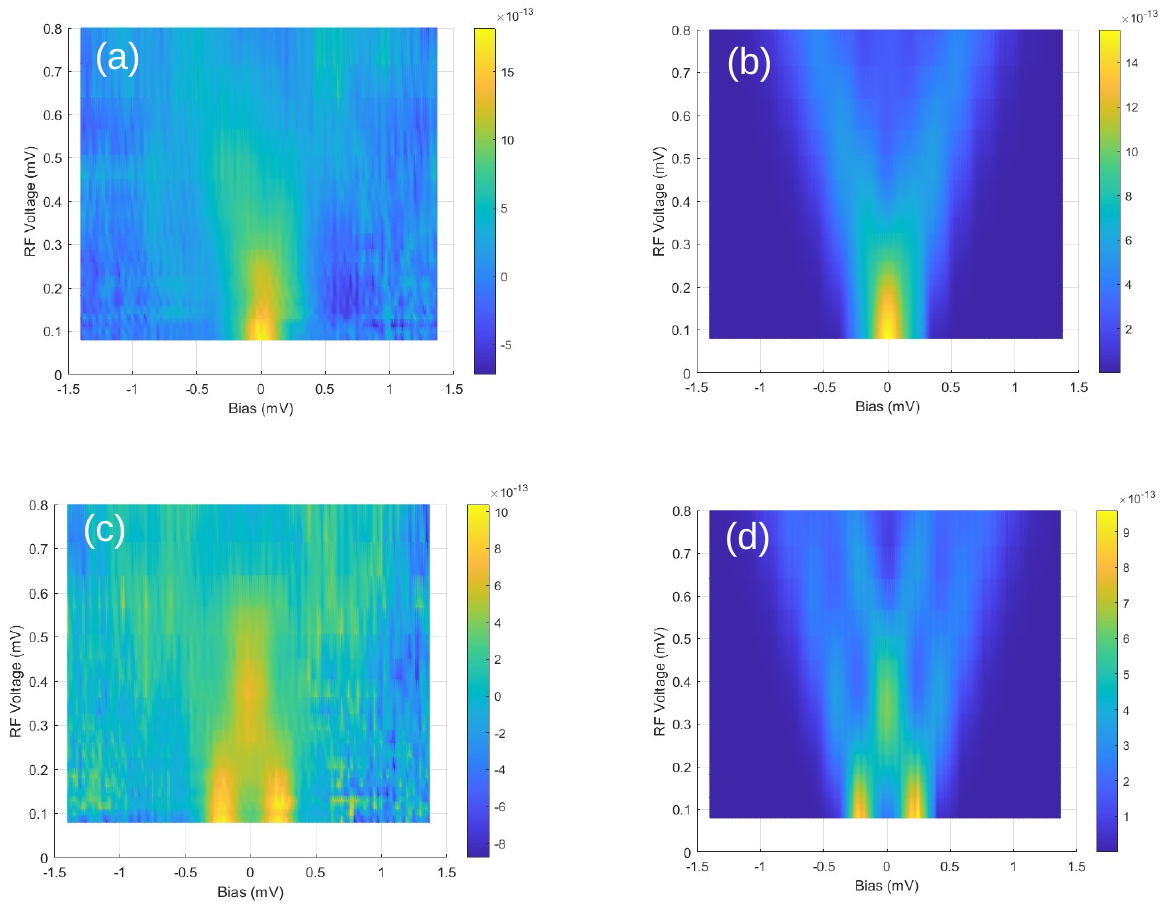}  
\caption{(a) Differential conductance (color plot) as a function of DC bias and the RF voltage, obtained on the third atom of the Cr$_{12}$ chain, see Suppl. Information for details. (b) Tien-Gordon simulation using the zero-RF signal for the third atom. These results show that the RF vs DC behavior is rather symmetrical with respect to the zero DC bias. (c) and (d) the same type of measurement and simulation for the sixth atom of the chain showing a clear topological gap matching the one at zero driving. As the $V_{RF}$ bias is ramped, the conductance increases zero DC bias in the middle of the chain.}
\label{figureS5}
\end{figure*}

An interesting approach is to study the zero-energy modes by radio-frequency (RF) excitation, following the studies where the in-gap states are excited by microwave radiation~\cite{peters_resonant_2020,gonzalez_photon-assisted_2020}. Figure~\ref{figureS5} shows the results of our measurements (a) and (c), together with our fitting (b) and (d) using the theory of Ref.~\cite{kot_microwave-assisted_2020}. The fittings clearly show the expected symmetric distribution of the $dI/dV$ peaks displaced by the microwave absorption. The experimental signal-to-noise ratio is lower than in previous microwave-assisted tunnelling studies, preventing a more precise determination of the residual offset of the edge mode from zero bias~\cite{peters_resonant_2020,kot_microwave-assisted_2020}.

Interestingly, we see that in the middle of chain (c) shows the clear gap and as the RF bias is increased a peak in the middle of the gap appearing. This new peak does not appear in the microwave study of Fig.~4 because it corresponds to much larger $V_{RF}$ values. The data of Fig.~4 correspond to the fuzzy area about $V_{RF} \sim 0.2$~mV, where indeed the microwave smears and broadens the signal as capture by our Dynes-broadening parameter of the main text, Fig. \ref{figureS5}. 

\begin{figure*}[t!]
\centering
\vspace{-4.cm}
\includegraphics[width=1.\linewidth]{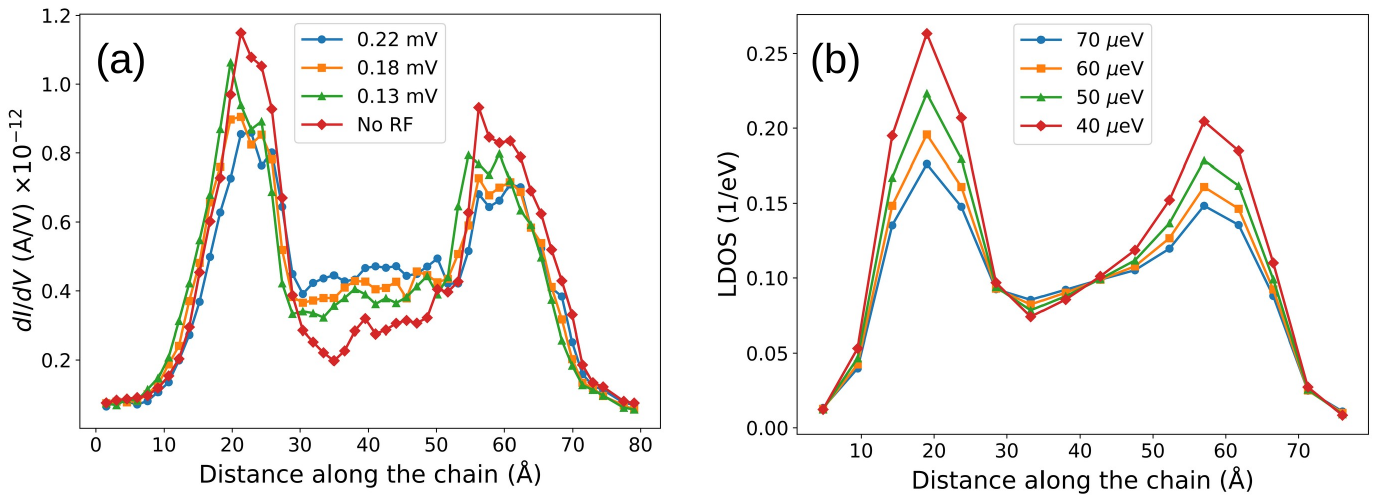}
\vspace{-1.5cm}
\caption{
(a) Zero-bias differential-conductance profiles measured along the Cr$_{12}$
chain for increasing microwave amplitudes. ``No RF'' denotes the absence of
radio-frequency driving, while $V_{\rm RF}=0.22$ mV corresponds to the largest
applied ac modulation. Increasing $V_{\rm RF}$ reduces the zero-bias
conductance at the chain ends and enhances the signal in the central region.
(b) Simulated zero-energy LDOS profiles obtained by increasing a Dynes
broadening parameter, as indicated in the legend. Larger values correspond to
stronger quasiparticle broadening. The agreement between experiment and
simulation indicates that GHz bias modulation acts as a dynamical broadening
probe of the compact zero-energy modes, redistributing spectral weight without
requiring a displacement of the mode away from zero bias. The sensitivity of
the profile to moderate broadening is consistent with the short
$\sim 3$-atom localization length extracted from the residual end-mode
hybridization.
}
\label{Sfigure5}
\end{figure*}
\clearpage

\section*{Acknowledgements}
We thank Mireia Tena and Cristina Mier for their support at the early stage of the experiment.
NL and DJC thank projects PID2021-127917NB-I00 by MCIN/AEI/10.13039/501100011033, QUAN-000021-01 by Gipuzkoa Provincial Council, PIBA-2024-1-0008 and IT-1527-22 by Basque Government, ESiM project 101046364 by EU. Views and opinions expressed are however those of the author(s) only and do not necessarily reflect those of the EU. Neither the EU nor the granting authority can be held responsible for them. JK, PK and AJH acknowledge support from Institute for Basic Science (grant: IBS-R027-D1).

\bibliography{references}

\begin{thebibliography}{41}%
\makeatletter
\providecommand \@ifxundefined [1]{%
 \@ifx{#1\undefined}
}%
\providecommand \@ifnum [1]{%
 \ifnum #1\expandafter \@firstoftwo
 \else \expandafter \@secondoftwo
 \fi
}%
\providecommand \@ifx [1]{%
 \ifx #1\expandafter \@firstoftwo
 \else \expandafter \@secondoftwo
 \fi
}%
\providecommand \natexlab [1]{#1}%
\providecommand \enquote  [1]{``#1''}%
\providecommand \bibnamefont  [1]{#1}%
\providecommand \bibfnamefont [1]{#1}%
\providecommand \citenamefont [1]{#1}%
\providecommand \href@noop [0]{\@secondoftwo}%
\providecommand \href [0]{\begingroup \@sanitize@url \@href}%
\providecommand \@href[1]{\@@startlink{#1}\@@href}%
\providecommand \@@href[1]{\endgroup#1\@@endlink}%
\providecommand \@sanitize@url [0]{\catcode `\\12\catcode `\$12\catcode
  `\&12\catcode `\#12\catcode `\^12\catcode `\_12\catcode `\%12\relax}%
\providecommand \@@startlink[1]{}%
\providecommand \@@endlink[0]{}%
\providecommand \url  [0]{\begingroup\@sanitize@url \@url }%
\providecommand \@url [1]{\endgroup\@href {#1}{\urlprefix }}%
\providecommand \urlprefix  [0]{URL }%
\providecommand \Eprint [0]{\href }%
\providecommand \doibase [0]{https://doi.org/}%
\providecommand \selectlanguage [0]{\@gobble}%
\providecommand \bibinfo  [0]{\@secondoftwo}%
\providecommand \bibfield  [0]{\@secondoftwo}%
\providecommand \translation [1]{[#1]}%
\providecommand \BibitemOpen [0]{}%
\providecommand \bibitemStop [0]{}%
\providecommand \bibitemNoStop [0]{.\EOS\space}%
\providecommand \EOS [0]{\spacefactor3000\relax}%
\providecommand \BibitemShut  [1]{\csname bibitem#1\endcsname}%
\let\auto@bib@innerbib\@empty
\bibitem [{\citenamefont {Yazdani}\ \emph {et~al.}(2023)\citenamefont
  {Yazdani}, \citenamefont {von Oppen}, \citenamefont {Halperin},\ and\
  \citenamefont {Yacoby}}]{Yazdani_2023}%
  \BibitemOpen
  \bibfield  {author} {\bibinfo {author} {\bibfnamefont {A.}~\bibnamefont
  {Yazdani}}, \bibinfo {author} {\bibfnamefont {F.}~\bibnamefont {von Oppen}},
  \bibinfo {author} {\bibfnamefont {B.~I.}\ \bibnamefont {Halperin}},\ and\
  \bibinfo {author} {\bibfnamefont {A.}~\bibnamefont {Yacoby}},\ }\bibfield
  {title} {\bibinfo {title} {Hunting for majoranas},\ }\href
  {https://doi.org/10.1126/science.ade0850} {\bibfield  {journal} {\bibinfo
  {journal} {Science}\ }\textbf {\bibinfo {volume} {380}},\ \bibinfo {pages}
  {eade0850} (\bibinfo {year} {2023})},\ \Eprint
  {https://arxiv.org/abs/https://www.science.org/doi/pdf/10.1126/science.ade0850}
  {https://www.science.org/doi/pdf/10.1126/science.ade0850} \BibitemShut
  {NoStop}%
\bibitem [{\citenamefont {Choy}\ \emph {et~al.}(2011)\citenamefont {Choy},
  \citenamefont {Edge}, \citenamefont {Akhmerov},\ and\ \citenamefont
  {Beenakker}}]{Beenakker}%
  \BibitemOpen
  \bibfield  {author} {\bibinfo {author} {\bibfnamefont {T.-P.}\ \bibnamefont
  {Choy}}, \bibinfo {author} {\bibfnamefont {J.~M.}\ \bibnamefont {Edge}},
  \bibinfo {author} {\bibfnamefont {A.~R.}\ \bibnamefont {Akhmerov}},\ and\
  \bibinfo {author} {\bibfnamefont {C.~W.~J.}\ \bibnamefont {Beenakker}},\
  }\bibfield  {title} {\bibinfo {title} {Majorana fermions emerging from
  magnetic nanoparticles on a superconductor without spin-orbit coupling},\
  }\href {https://doi.org/10.1103/PhysRevB.84.195442} {\bibfield  {journal}
  {\bibinfo  {journal} {Physical Review B}\ }\textbf {\bibinfo {volume} {84}},\
  \bibinfo {pages} {195442} (\bibinfo {year} {2011})}\BibitemShut {NoStop}%
\bibitem [{\citenamefont {Nadj-Perge}\ \emph {et~al.}(2013)\citenamefont
  {Nadj-Perge}, \citenamefont {Drozdov}, \citenamefont {Bernevig},\ and\
  \citenamefont {Yazdani}}]{Yazdani_2013}%
  \BibitemOpen
  \bibfield  {author} {\bibinfo {author} {\bibfnamefont {S.}~\bibnamefont
  {Nadj-Perge}}, \bibinfo {author} {\bibfnamefont {I.~K.}\ \bibnamefont
  {Drozdov}}, \bibinfo {author} {\bibfnamefont {B.~A.}\ \bibnamefont
  {Bernevig}},\ and\ \bibinfo {author} {\bibfnamefont {A.}~\bibnamefont
  {Yazdani}},\ }\bibfield  {title} {\bibinfo {title} {Proposal for realizing
  majorana fermions in chains of magnetic atoms on a superconductor},\ }\href
  {https://doi.org/10.1103/PhysRevB.88.020407} {\bibfield  {journal} {\bibinfo
  {journal} {Phys. Rev. B}\ }\textbf {\bibinfo {volume} {88}},\ \bibinfo
  {pages} {020407} (\bibinfo {year} {2013})}\BibitemShut {NoStop}%
\bibitem [{\citenamefont {Nadj-Perge}\ \emph {et~al.}(2014)\citenamefont
  {Nadj-Perge}, \citenamefont {Drozdov}, \citenamefont {Li}, \citenamefont
  {Chen}, \citenamefont {Jeon}, \citenamefont {Seo}, \citenamefont {MacDonald},
  \citenamefont {Bernevig},\ and\ \citenamefont {Yazdani}}]{Yazdani}%
  \BibitemOpen
  \bibfield  {author} {\bibinfo {author} {\bibfnamefont {S.}~\bibnamefont
  {Nadj-Perge}}, \bibinfo {author} {\bibfnamefont {I.~K.}\ \bibnamefont
  {Drozdov}}, \bibinfo {author} {\bibfnamefont {J.}~\bibnamefont {Li}},
  \bibinfo {author} {\bibfnamefont {H.}~\bibnamefont {Chen}}, \bibinfo {author}
  {\bibfnamefont {S.}~\bibnamefont {Jeon}}, \bibinfo {author} {\bibfnamefont
  {J.}~\bibnamefont {Seo}}, \bibinfo {author} {\bibfnamefont {A.~H.}\
  \bibnamefont {MacDonald}}, \bibinfo {author} {\bibfnamefont {B.~A.}\
  \bibnamefont {Bernevig}},\ and\ \bibinfo {author} {\bibfnamefont
  {A.}~\bibnamefont {Yazdani}},\ }\bibfield  {title} {\bibinfo {title}
  {Observation of {Majorana} fermions in ferromagnetic atomic chains on a
  superconductor},\ }\href {https://doi.org/10.1126/science.1259327} {\bibfield
   {journal} {\bibinfo  {journal} {Science}\ }\textbf {\bibinfo {volume}
  {346}},\ \bibinfo {pages} {602} (\bibinfo {year} {2014})}\BibitemShut
  {NoStop}%
\bibitem [{\citenamefont {Ruby}\ \emph {et~al.}(2015)\citenamefont {Ruby},
  \citenamefont {Pientka}, \citenamefont {Peng}, \citenamefont {von Oppen},
  \citenamefont {Heinrich},\ and\ \citenamefont {Franke}}]{Ruby_chains}%
  \BibitemOpen
  \bibfield  {author} {\bibinfo {author} {\bibfnamefont {M.}~\bibnamefont
  {Ruby}}, \bibinfo {author} {\bibfnamefont {F.}~\bibnamefont {Pientka}},
  \bibinfo {author} {\bibfnamefont {Y.}~\bibnamefont {Peng}}, \bibinfo {author}
  {\bibfnamefont {F.}~\bibnamefont {von Oppen}}, \bibinfo {author}
  {\bibfnamefont {B.~W.}\ \bibnamefont {Heinrich}},\ and\ \bibinfo {author}
  {\bibfnamefont {K.~J.}\ \bibnamefont {Franke}},\ }\bibfield  {title}
  {\bibinfo {title} {End states and subgap structure in proximity-coupled
  chains of magnetic adatoms},\ }\href
  {https://doi.org/10.1103/PhysRevLett.115.197204} {\bibfield  {journal}
  {\bibinfo  {journal} {Phys. Rev. Lett.}\ }\textbf {\bibinfo {volume} {115}},\
  \bibinfo {pages} {197204} (\bibinfo {year} {2015})}\BibitemShut {NoStop}%
\bibitem [{\citenamefont {Pawlak}\ \emph {et~al.}(2016)\citenamefont {Pawlak},
  \citenamefont {Kisiel}, \citenamefont {Klinovaja}, \citenamefont {Meier},
  \citenamefont {Kawai}, \citenamefont {Glatzel}, \citenamefont {Loss},\ and\
  \citenamefont {Meyer}}]{Pawlak_2016}%
  \BibitemOpen
  \bibfield  {author} {\bibinfo {author} {\bibfnamefont {R.}~\bibnamefont
  {Pawlak}}, \bibinfo {author} {\bibfnamefont {M.}~\bibnamefont {Kisiel}},
  \bibinfo {author} {\bibfnamefont {J.}~\bibnamefont {Klinovaja}}, \bibinfo
  {author} {\bibfnamefont {T.}~\bibnamefont {Meier}}, \bibinfo {author}
  {\bibfnamefont {S.}~\bibnamefont {Kawai}}, \bibinfo {author} {\bibfnamefont
  {T.}~\bibnamefont {Glatzel}}, \bibinfo {author} {\bibfnamefont
  {D.}~\bibnamefont {Loss}},\ and\ \bibinfo {author} {\bibfnamefont
  {E.}~\bibnamefont {Meyer}},\ }\bibfield  {title} {\bibinfo {title} {Probing
  atomic structure and majorana wavefunctions in mono-atomic fe chains on
  superconducting pb surface},\ }\href@noop {} {\bibfield  {journal} {\bibinfo
  {journal} {Npj Quantum Information}\ }\textbf {\bibinfo {volume} {2}},\
  \bibinfo {pages} {16035} (\bibinfo {year} {2016})}\BibitemShut {NoStop}%
\bibitem [{\citenamefont {Kim}\ \emph {et~al.}(2018)\citenamefont {Kim},
  \citenamefont {Palacio-Morales}, \citenamefont {Posske}, \citenamefont
  {R{\'o}zsa}, \citenamefont {Palot{\'a}s}, \citenamefont {Szunyogh},
  \citenamefont {Thorwart},\ and\ \citenamefont {Wiesendanger}}]{Kim_2018}%
  \BibitemOpen
  \bibfield  {author} {\bibinfo {author} {\bibfnamefont {H.}~\bibnamefont
  {Kim}}, \bibinfo {author} {\bibfnamefont {A.}~\bibnamefont
  {Palacio-Morales}}, \bibinfo {author} {\bibfnamefont {T.}~\bibnamefont
  {Posske}}, \bibinfo {author} {\bibfnamefont {L.}~\bibnamefont {R{\'o}zsa}},
  \bibinfo {author} {\bibfnamefont {K.}~\bibnamefont {Palot{\'a}s}}, \bibinfo
  {author} {\bibfnamefont {L.}~\bibnamefont {Szunyogh}}, \bibinfo {author}
  {\bibfnamefont {M.}~\bibnamefont {Thorwart}},\ and\ \bibinfo {author}
  {\bibfnamefont {R.}~\bibnamefont {Wiesendanger}},\ }\bibfield  {title}
  {\bibinfo {title} {Toward tailoring majorana bound states in artificially
  constructed magnetic atom chains on elemental superconductors},\ }\href
  {https://doi.org/10.1126/sciadv.aar5251} {\bibfield  {journal} {\bibinfo
  {journal} {Science Advances}\ }\textbf {\bibinfo {volume} {4}},\ \bibinfo
  {pages} {eaar5251} (\bibinfo {year} {2018})}\BibitemShut {NoStop}%
\bibitem [{\citenamefont {Li}\ \emph {et~al.}(2018)\citenamefont {Li},
  \citenamefont {Jeon}, \citenamefont {Xie}, \citenamefont {Yazdani},\ and\
  \citenamefont {Bernevig}}]{Li_2018}%
  \BibitemOpen
  \bibfield  {author} {\bibinfo {author} {\bibfnamefont {J.}~\bibnamefont
  {Li}}, \bibinfo {author} {\bibfnamefont {S.}~\bibnamefont {Jeon}}, \bibinfo
  {author} {\bibfnamefont {Y.}~\bibnamefont {Xie}}, \bibinfo {author}
  {\bibfnamefont {A.}~\bibnamefont {Yazdani}},\ and\ \bibinfo {author}
  {\bibfnamefont {B.~A.}\ \bibnamefont {Bernevig}},\ }\bibfield  {title}
  {\bibinfo {title} {Majorana spin in magnetic atomic chain systems},\
  }\bibfield  {journal} {\bibinfo  {journal} {Physical Review B}\ }\textbf
  {\bibinfo {volume} {97}},\ \href {https://doi.org/10.1103/PhysRevB.97.125119}
  {10.1103/PhysRevB.97.125119} (\bibinfo {year} {2018})\BibitemShut {NoStop}%
\bibitem [{\citenamefont {Choi}\ \emph {et~al.}(2019)\citenamefont {Choi},
  \citenamefont {Lorente}, \citenamefont {Wiebe}, \citenamefont {von Bergmann},
  \citenamefont {Otte},\ and\ \citenamefont {Heinrich}}]{Choi_2019}%
  \BibitemOpen
  \bibfield  {author} {\bibinfo {author} {\bibfnamefont {D.-J.}\ \bibnamefont
  {Choi}}, \bibinfo {author} {\bibfnamefont {N.}~\bibnamefont {Lorente}},
  \bibinfo {author} {\bibfnamefont {J.}~\bibnamefont {Wiebe}}, \bibinfo
  {author} {\bibfnamefont {K.}~\bibnamefont {von Bergmann}}, \bibinfo {author}
  {\bibfnamefont {A.~F.}\ \bibnamefont {Otte}},\ and\ \bibinfo {author}
  {\bibfnamefont {A.~J.}\ \bibnamefont {Heinrich}},\ }\bibfield  {title}
  {\bibinfo {title} {Colloquium: {Atomic} spin chains on surfaces},\ }\href
  {https://doi.org/10.1103/RevModPhys.91.041001} {\bibfield  {journal}
  {\bibinfo  {journal} {Reviews of Modern Physics}\ }\textbf {\bibinfo {volume}
  {91}},\ \bibinfo {pages} {041001} (\bibinfo {year} {2019})}\BibitemShut
  {NoStop}%
\bibitem [{\citenamefont {Schneider}\ \emph
  {et~al.}(2021{\natexlab{a}})\citenamefont {Schneider}, \citenamefont {Beck},
  \citenamefont {Posske}, \citenamefont {Crawford}, \citenamefont {Mascot},
  \citenamefont {Rachel}, \citenamefont {Wiesendanger},\ and\ \citenamefont
  {Wiebe}}]{Schneider1}%
  \BibitemOpen
  \bibfield  {author} {\bibinfo {author} {\bibfnamefont {L.}~\bibnamefont
  {Schneider}}, \bibinfo {author} {\bibfnamefont {P.}~\bibnamefont {Beck}},
  \bibinfo {author} {\bibfnamefont {T.}~\bibnamefont {Posske}}, \bibinfo
  {author} {\bibfnamefont {D.}~\bibnamefont {Crawford}}, \bibinfo {author}
  {\bibfnamefont {E.}~\bibnamefont {Mascot}}, \bibinfo {author} {\bibfnamefont
  {S.}~\bibnamefont {Rachel}}, \bibinfo {author} {\bibfnamefont
  {R.}~\bibnamefont {Wiesendanger}},\ and\ \bibinfo {author} {\bibfnamefont
  {J.}~\bibnamefont {Wiebe}},\ }\bibfield  {title} {\bibinfo {title}
  {Topological {Shiba} bands in artificial spin chains on superconductors},\
  }\href {https://doi.org/10.1038/s41567-021-01234-y} {\bibfield  {journal}
  {\bibinfo  {journal} {Nature Physics}\ }\textbf {\bibinfo {volume} {17}},\
  \bibinfo {pages} {943 } (\bibinfo {year} {2021}{\natexlab{a}})}\BibitemShut
  {NoStop}%
\bibitem [{\citenamefont {Schneider}\ \emph {et~al.}(2022)\citenamefont
  {Schneider}, \citenamefont {Beck}, \citenamefont {Neuhaus-Steinmetz},
  \citenamefont {Rózsa}, \citenamefont {Posske}, \citenamefont {Wiebe},\ and\
  \citenamefont {Wiesendanger}}]{schneider_precursors_2022}%
  \BibitemOpen
  \bibfield  {author} {\bibinfo {author} {\bibfnamefont {L.}~\bibnamefont
  {Schneider}}, \bibinfo {author} {\bibfnamefont {P.}~\bibnamefont {Beck}},
  \bibinfo {author} {\bibfnamefont {J.}~\bibnamefont {Neuhaus-Steinmetz}},
  \bibinfo {author} {\bibfnamefont {L.}~\bibnamefont {Rózsa}}, \bibinfo
  {author} {\bibfnamefont {T.}~\bibnamefont {Posske}}, \bibinfo {author}
  {\bibfnamefont {J.}~\bibnamefont {Wiebe}},\ and\ \bibinfo {author}
  {\bibfnamefont {R.}~\bibnamefont {Wiesendanger}},\ }\bibfield  {title}
  {\bibinfo {title} {Precursors of {Majorana} modes and their length-dependent
  energy oscillations probed at both ends of atomic {Shiba} chains},\ }\href
  {https://doi.org/10.1038/s41565-022-01078-4} {\bibfield  {journal} {\bibinfo
  {journal} {Nature Nanotechnology}\ }\textbf {\bibinfo {volume} {17}},\
  \bibinfo {pages} {384} (\bibinfo {year} {2022})}\BibitemShut {NoStop}%
\bibitem [{\citenamefont {Beck}\ \emph {et~al.}(2023)\citenamefont {Beck},
  \citenamefont {Ny{\'a}ri}, \citenamefont {Schneider}, \citenamefont
  {R{\'o}zsa}, \citenamefont {L{\'a}szl{\'o}ffy}, \citenamefont {Palot{\'a}s},
  \citenamefont {Szunyogh}, \citenamefont {Ujfalussy}, \citenamefont {Wiebe},\
  and\ \citenamefont {Wiesendanger}}]{Beck2023_large_topological_gaps}%
  \BibitemOpen
  \bibfield  {author} {\bibinfo {author} {\bibfnamefont {P.}~\bibnamefont
  {Beck}}, \bibinfo {author} {\bibfnamefont {B.}~\bibnamefont {Ny{\'a}ri}},
  \bibinfo {author} {\bibfnamefont {L.}~\bibnamefont {Schneider}}, \bibinfo
  {author} {\bibfnamefont {L.}~\bibnamefont {R{\'o}zsa}}, \bibinfo {author}
  {\bibfnamefont {A.}~\bibnamefont {L{\'a}szl{\'o}ffy}}, \bibinfo {author}
  {\bibfnamefont {K.}~\bibnamefont {Palot{\'a}s}}, \bibinfo {author}
  {\bibfnamefont {L.}~\bibnamefont {Szunyogh}}, \bibinfo {author}
  {\bibfnamefont {B.}~\bibnamefont {Ujfalussy}}, \bibinfo {author}
  {\bibfnamefont {J.}~\bibnamefont {Wiebe}},\ and\ \bibinfo {author}
  {\bibfnamefont {R.}~\bibnamefont {Wiesendanger}},\ }\bibfield  {title}
  {\bibinfo {title} {Search for large topological gaps in atomic spin chains on
  proximitized superconducting heavy metal layers},\ }\href
  {https://doi.org/10.1038/s42005-023-01196-y} {\bibfield  {journal} {\bibinfo
  {journal} {Communications Physics}\ }\textbf {\bibinfo {volume} {6}},\
  \bibinfo {pages} {83} (\bibinfo {year} {2023})}\BibitemShut {NoStop}%
\bibitem [{\citenamefont {Lo~Conte}\ \emph {et~al.}(2024)\citenamefont
  {Lo~Conte}, \citenamefont {Wiebe}, \citenamefont {Rachel}, \citenamefont
  {Morr},\ and\ \citenamefont
  {Wiesendanger}}]{lo_conte_magnet-superconductor_2024}%
  \BibitemOpen
  \bibfield  {author} {\bibinfo {author} {\bibfnamefont {R.}~\bibnamefont
  {Lo~Conte}}, \bibinfo {author} {\bibfnamefont {J.}~\bibnamefont {Wiebe}},
  \bibinfo {author} {\bibfnamefont {S.}~\bibnamefont {Rachel}}, \bibinfo
  {author} {\bibfnamefont {D.~K.}\ \bibnamefont {Morr}},\ and\ \bibinfo
  {author} {\bibfnamefont {R.}~\bibnamefont {Wiesendanger}},\ }\bibfield
  {title} {\bibinfo {title} {Magnet-superconductor hybrid quantum systems: a
  materials platform for topological superconductivity},\ }\href
  {https://doi.org/10.1007/s40766-024-00060-1} {\bibfield  {journal} {\bibinfo
  {journal} {La Rivista del Nuovo Cimento}\ }\textbf {\bibinfo {volume} {47}},\
  \bibinfo {pages} {453} (\bibinfo {year} {2024})}\BibitemShut {NoStop}%
\bibitem [{\citenamefont {Yu}\ \emph {et~al.}(2021)\citenamefont {Yu},
  \citenamefont {Chen}, \citenamefont {Gomanko}, \citenamefont {Badawy},
  \citenamefont {Bakkers}, \citenamefont {Zuo}, \citenamefont {Mourik},\ and\
  \citenamefont {Frolov}}]{Frolov_2021}%
  \BibitemOpen
  \bibfield  {author} {\bibinfo {author} {\bibfnamefont {P.}~\bibnamefont
  {Yu}}, \bibinfo {author} {\bibfnamefont {J.}~\bibnamefont {Chen}}, \bibinfo
  {author} {\bibfnamefont {M.}~\bibnamefont {Gomanko}}, \bibinfo {author}
  {\bibfnamefont {G.}~\bibnamefont {Badawy}}, \bibinfo {author} {\bibfnamefont
  {E.~P. A.~M.}\ \bibnamefont {Bakkers}}, \bibinfo {author} {\bibfnamefont
  {K.}~\bibnamefont {Zuo}}, \bibinfo {author} {\bibfnamefont {V.}~\bibnamefont
  {Mourik}},\ and\ \bibinfo {author} {\bibfnamefont {S.~M.}\ \bibnamefont
  {Frolov}},\ }\bibfield  {title} {\bibinfo {title} {Non-{Majorana} states
  yield nearly quantized conductance in proximatized nanowires},\ }\href
  {https://doi.org/10.1038/s41567-020-01107-w} {\bibfield  {journal} {\bibinfo
  {journal} {Nature Physics}\ }\textbf {\bibinfo {volume} {17}},\ \bibinfo
  {pages} {482} (\bibinfo {year} {2021})}\BibitemShut {NoStop}%
\bibitem [{\citenamefont {Dmytruk}\ \emph {et~al.}(2020)\citenamefont
  {Dmytruk}, \citenamefont {Loss},\ and\ \citenamefont
  {Klinovaja}}]{PhysRevB.102.245431}%
  \BibitemOpen
  \bibfield  {author} {\bibinfo {author} {\bibfnamefont {O.}~\bibnamefont
  {Dmytruk}}, \bibinfo {author} {\bibfnamefont {D.}~\bibnamefont {Loss}},\ and\
  \bibinfo {author} {\bibfnamefont {J.}~\bibnamefont {Klinovaja}},\ }\bibfield
  {title} {\bibinfo {title} {Pinning of andreev bound states to zero energy in
  two-dimensional superconductor- semiconductor rashba heterostructures},\
  }\href {https://doi.org/10.1103/PhysRevB.102.245431} {\bibfield  {journal}
  {\bibinfo  {journal} {Phys. Rev. B}\ }\textbf {\bibinfo {volume} {102}},\
  \bibinfo {pages} {245431} (\bibinfo {year} {2020})}\BibitemShut {NoStop}%
\bibitem [{\citenamefont {Hess}\ \emph {et~al.}(2022)\citenamefont {Hess},
  \citenamefont {Legg}, \citenamefont {Loss},\ and\ \citenamefont
  {Klinovaja}}]{PhysRevB.106.104503}%
  \BibitemOpen
  \bibfield  {author} {\bibinfo {author} {\bibfnamefont {R.}~\bibnamefont
  {Hess}}, \bibinfo {author} {\bibfnamefont {H.~F.}\ \bibnamefont {Legg}},
  \bibinfo {author} {\bibfnamefont {D.}~\bibnamefont {Loss}},\ and\ \bibinfo
  {author} {\bibfnamefont {J.}~\bibnamefont {Klinovaja}},\ }\bibfield  {title}
  {\bibinfo {title} {Prevalence of trivial zero-energy subgap states in
  nonuniform helical spin chains on the surface of superconductors},\ }\href
  {https://doi.org/10.1103/PhysRevB.106.104503} {\bibfield  {journal} {\bibinfo
   {journal} {Phys. Rev. B}\ }\textbf {\bibinfo {volume} {106}},\ \bibinfo
  {pages} {104503} (\bibinfo {year} {2022})}\BibitemShut {NoStop}%
\bibitem [{\citenamefont {Küster}\ \emph {et~al.}(2022)\citenamefont
  {Küster}, \citenamefont {Brinker}, \citenamefont {Hess}, \citenamefont
  {Loss}, \citenamefont {Parkin}, \citenamefont {Klinovaja}, \citenamefont
  {Lounis},\ and\ \citenamefont {Sessi}}]{kuster_non-majorana_2022}%
  \BibitemOpen
  \bibfield  {author} {\bibinfo {author} {\bibfnamefont {F.}~\bibnamefont
  {Küster}}, \bibinfo {author} {\bibfnamefont {S.}~\bibnamefont {Brinker}},
  \bibinfo {author} {\bibfnamefont {R.}~\bibnamefont {Hess}}, \bibinfo {author}
  {\bibfnamefont {D.}~\bibnamefont {Loss}}, \bibinfo {author} {\bibfnamefont
  {S.~S.~P.}\ \bibnamefont {Parkin}}, \bibinfo {author} {\bibfnamefont
  {J.}~\bibnamefont {Klinovaja}}, \bibinfo {author} {\bibfnamefont
  {S.}~\bibnamefont {Lounis}},\ and\ \bibinfo {author} {\bibfnamefont
  {P.}~\bibnamefont {Sessi}},\ }\bibfield  {title} {\bibinfo {title}
  {Non-{Majorana} modes in diluted spin chains proximitized to a
  superconductor},\ }\href {https://doi.org/10.1073/pnas.2210589119} {\bibfield
   {journal} {\bibinfo  {journal} {Proceedings of the National Academy of
  Sciences}\ }\textbf {\bibinfo {volume} {119}},\ \bibinfo {pages}
  {e2210589119} (\bibinfo {year} {2022})}\BibitemShut {NoStop}%
\bibitem [{\citenamefont {Schneider}\ \emph {et~al.}(2023)\citenamefont
  {Schneider}, \citenamefont {Beck}, \citenamefont {Rózsa}, \citenamefont
  {Posske}, \citenamefont {Wiebe},\ and\ \citenamefont
  {Wiesendanger}}]{schneider_probing_2023}%
  \BibitemOpen
  \bibfield  {author} {\bibinfo {author} {\bibfnamefont {L.}~\bibnamefont
  {Schneider}}, \bibinfo {author} {\bibfnamefont {P.}~\bibnamefont {Beck}},
  \bibinfo {author} {\bibfnamefont {L.}~\bibnamefont {Rózsa}}, \bibinfo
  {author} {\bibfnamefont {T.}~\bibnamefont {Posske}}, \bibinfo {author}
  {\bibfnamefont {J.}~\bibnamefont {Wiebe}},\ and\ \bibinfo {author}
  {\bibfnamefont {R.}~\bibnamefont {Wiesendanger}},\ }\bibfield  {title}
  {\bibinfo {title} {Probing the topologically trivial nature of end states in
  antiferromagnetic atomic chains on superconductors},\ }\href
  {https://doi.org/10.1038/s41467-023-38369-w} {\bibfield  {journal} {\bibinfo
  {journal} {Nature Communications}\ }\textbf {\bibinfo {volume} {14}},\
  \bibinfo {pages} {2742} (\bibinfo {year} {2023})}\BibitemShut {NoStop}%
\bibitem [{\citenamefont {Aguado}(2017)}]{Aguado}%
  \BibitemOpen
  \bibfield  {author} {\bibinfo {author} {\bibfnamefont {R.}~\bibnamefont
  {Aguado}},\ }\bibfield  {title} {\bibinfo {title} {Majorana quasiparticles in
  condensed matter},\ }\href {https://doi.org/10.1393/ncr/i2017-10141-9}
  {\bibfield  {journal} {\bibinfo  {journal} {La Rivista del Nuovo Cimento}\
  }\textbf {\bibinfo {volume} {40}},\ \bibinfo {pages} {523} (\bibinfo {year}
  {2017})}\BibitemShut {NoStop}%
\bibitem [{\citenamefont {Mier}\ \emph {et~al.}(2022)\citenamefont {Mier},
  \citenamefont {Choi},\ and\ \citenamefont {Lorente}}]{Choi2022}%
  \BibitemOpen
  \bibfield  {author} {\bibinfo {author} {\bibfnamefont {C.}~\bibnamefont
  {Mier}}, \bibinfo {author} {\bibfnamefont {D.-J.}\ \bibnamefont {Choi}},\
  and\ \bibinfo {author} {\bibfnamefont {N.}~\bibnamefont {Lorente}},\
  }\bibfield  {title} {\bibinfo {title} {Moir\'e dispersion of edge states in
  spin chains on superconductors},\ }\href
  {https://doi.org/10.1103/PhysRevResearch.4.L032010} {\bibfield  {journal}
  {\bibinfo  {journal} {Phys. Rev. Res.}\ }\textbf {\bibinfo {volume} {4}},\
  \bibinfo {pages} {L032010} (\bibinfo {year} {2022})}\BibitemShut {NoStop}%
\bibitem [{\citenamefont {Imai}\ \emph {et~al.}(2012)\citenamefont {Imai},
  \citenamefont {Nabeshima}, \citenamefont {Yoshinaka}, \citenamefont
  {Miyatani}, \citenamefont {Kondo}, \citenamefont {Komiya}, \citenamefont
  {Tsukada},\ and\ \citenamefont {Maeda}}]{imai_2012}%
  \BibitemOpen
  \bibfield  {author} {\bibinfo {author} {\bibfnamefont {Y.}~\bibnamefont
  {Imai}}, \bibinfo {author} {\bibfnamefont {F.}~\bibnamefont {Nabeshima}},
  \bibinfo {author} {\bibfnamefont {T.}~\bibnamefont {Yoshinaka}}, \bibinfo
  {author} {\bibfnamefont {K.}~\bibnamefont {Miyatani}}, \bibinfo {author}
  {\bibfnamefont {R.}~\bibnamefont {Kondo}}, \bibinfo {author} {\bibfnamefont
  {S.}~\bibnamefont {Komiya}}, \bibinfo {author} {\bibfnamefont
  {I.}~\bibnamefont {Tsukada}},\ and\ \bibinfo {author} {\bibfnamefont
  {A.}~\bibnamefont {Maeda}},\ }\bibfield  {title} {\bibinfo {title}
  {Superconductivity at 5.4 {K} in $\beta$-{Bi}$_2${Pd}},\ }\href
  {https://doi.org/10.1143/JPSJ.81.113708} {\bibfield  {journal} {\bibinfo
  {journal} {Journal of the Physical Society of Japan}\ }\textbf {\bibinfo
  {volume} {81}},\ \bibinfo {pages} {113708} (\bibinfo {year}
  {2012})}\BibitemShut {NoStop}%
\bibitem [{\citenamefont {Choi}\ \emph {et~al.}(2018)\citenamefont {Choi},
  \citenamefont {Fern\'andez}, \citenamefont {Herrera}, \citenamefont
  {Rubio-Verd\'u}, \citenamefont {Ugeda}, \citenamefont {Guillam\'on},
  \citenamefont {Suderow}, \citenamefont {Pascual},\ and\ \citenamefont
  {Lorente}}]{Choi_2018}%
  \BibitemOpen
  \bibfield  {author} {\bibinfo {author} {\bibfnamefont {D.-J.}\ \bibnamefont
  {Choi}}, \bibinfo {author} {\bibfnamefont {C.~G.}\ \bibnamefont
  {Fern\'andez}}, \bibinfo {author} {\bibfnamefont {E.}~\bibnamefont
  {Herrera}}, \bibinfo {author} {\bibfnamefont {C.}~\bibnamefont
  {Rubio-Verd\'u}}, \bibinfo {author} {\bibfnamefont {M.~M.}\ \bibnamefont
  {Ugeda}}, \bibinfo {author} {\bibfnamefont {I.}~\bibnamefont {Guillam\'on}},
  \bibinfo {author} {\bibfnamefont {H.}~\bibnamefont {Suderow}}, \bibinfo
  {author} {\bibfnamefont {J.~I.}\ \bibnamefont {Pascual}},\ and\ \bibinfo
  {author} {\bibfnamefont {N.}~\bibnamefont {Lorente}},\ }\bibfield  {title}
  {\bibinfo {title} {Influence of magnetic ordering between cr adatoms on the
  yu-shiba-rusinov states of the
  $\ensuremath{\beta}\text{\ensuremath{-}}{\mathrm{bi}}_{2}\mathrm{Pd}$
  superconductor},\ }\href {https://doi.org/10.1103/PhysRevLett.120.167001}
  {\bibfield  {journal} {\bibinfo  {journal} {Phys. Rev. Lett.}\ }\textbf
  {\bibinfo {volume} {120}},\ \bibinfo {pages} {167001} (\bibinfo {year}
  {2018})}\BibitemShut {NoStop}%
\bibitem [{\citenamefont {Mier}\ \emph
  {et~al.}(2021{\natexlab{a}})\citenamefont {Mier}, \citenamefont {Hwang},
  \citenamefont {Kim}, \citenamefont {Bae}, \citenamefont {Nabeshima},
  \citenamefont {Imai}, \citenamefont {Maeda}, \citenamefont {Lorente},
  \citenamefont {Heinrich},\ and\ \citenamefont {Choi}}]{Choi2021}%
  \BibitemOpen
  \bibfield  {author} {\bibinfo {author} {\bibfnamefont {C.}~\bibnamefont
  {Mier}}, \bibinfo {author} {\bibfnamefont {J.}~\bibnamefont {Hwang}},
  \bibinfo {author} {\bibfnamefont {J.}~\bibnamefont {Kim}}, \bibinfo {author}
  {\bibfnamefont {Y.}~\bibnamefont {Bae}}, \bibinfo {author} {\bibfnamefont
  {F.}~\bibnamefont {Nabeshima}}, \bibinfo {author} {\bibfnamefont
  {Y.}~\bibnamefont {Imai}}, \bibinfo {author} {\bibfnamefont {A.}~\bibnamefont
  {Maeda}}, \bibinfo {author} {\bibfnamefont {N.}~\bibnamefont {Lorente}},
  \bibinfo {author} {\bibfnamefont {A.}~\bibnamefont {Heinrich}},\ and\
  \bibinfo {author} {\bibfnamefont {D.-J.}\ \bibnamefont {Choi}},\ }\bibfield
  {title} {\bibinfo {title} {Atomic manipulation of in-gap states in the
  $\ensuremath{\beta}\mathrm{-Bi}_{2}\mathrm{Pd}$ superconductor},\ }\href
  {https://doi.org/10.1103/PhysRevB.104.045406} {\bibfield  {journal} {\bibinfo
   {journal} {Phys. Rev. B}\ }\textbf {\bibinfo {volume} {104}},\ \bibinfo
  {pages} {045406} (\bibinfo {year} {2021}{\natexlab{a}})}\BibitemShut
  {NoStop}%
\bibitem [{\citenamefont {Peng}\ \emph {et~al.}(2015)\citenamefont {Peng},
  \citenamefont {Pientka}, \citenamefont {Glazman},\ and\ \citenamefont {von
  Oppen}}]{Peng_2015}%
  \BibitemOpen
  \bibfield  {author} {\bibinfo {author} {\bibfnamefont {Y.}~\bibnamefont
  {Peng}}, \bibinfo {author} {\bibfnamefont {F.}~\bibnamefont {Pientka}},
  \bibinfo {author} {\bibfnamefont {L.~I.}\ \bibnamefont {Glazman}},\ and\
  \bibinfo {author} {\bibfnamefont {F.}~\bibnamefont {von Oppen}},\ }\bibfield
  {title} {\bibinfo {title} {Strong localization of majorana end states in
  chains of magnetic adatoms},\ }\href
  {https://doi.org/10.1103/PhysRevLett.114.106801} {\bibfield  {journal}
  {\bibinfo  {journal} {Phys. Rev. Lett.}\ }\textbf {\bibinfo {volume} {114}},\
  \bibinfo {pages} {106801} (\bibinfo {year} {2015})}\BibitemShut {NoStop}%
\bibitem [{\citenamefont {Schecter}\ \emph {et~al.}(2016)\citenamefont
  {Schecter}, \citenamefont {Flensberg}, \citenamefont {Christensen},
  \citenamefont {Andersen},\ and\ \citenamefont {Paaske}}]{Paaske_2016}%
  \BibitemOpen
  \bibfield  {author} {\bibinfo {author} {\bibfnamefont {M.}~\bibnamefont
  {Schecter}}, \bibinfo {author} {\bibfnamefont {K.}~\bibnamefont {Flensberg}},
  \bibinfo {author} {\bibfnamefont {M.~H.}\ \bibnamefont {Christensen}},
  \bibinfo {author} {\bibfnamefont {B.~M.}\ \bibnamefont {Andersen}},\ and\
  \bibinfo {author} {\bibfnamefont {J.}~\bibnamefont {Paaske}},\ }\bibfield
  {title} {\bibinfo {title} {Self-organized topological superconductivity in a
  yu-shiba-rusinov chain},\ }\href {https://doi.org/10.1103/PhysRevB.93.140503}
  {\bibfield  {journal} {\bibinfo  {journal} {Phys. Rev. B}\ }\textbf {\bibinfo
  {volume} {93}},\ \bibinfo {pages} {140503} (\bibinfo {year}
  {2016})}\BibitemShut {NoStop}%
\bibitem [{\citenamefont {Mier}\ \emph
  {et~al.}(2021{\natexlab{b}})\citenamefont {Mier}, \citenamefont {Choi},\ and\
  \citenamefont {Lorente}}]{Mier2021_b}%
  \BibitemOpen
  \bibfield  {author} {\bibinfo {author} {\bibfnamefont {C.}~\bibnamefont
  {Mier}}, \bibinfo {author} {\bibfnamefont {D.-J.}\ \bibnamefont {Choi}},\
  and\ \bibinfo {author} {\bibfnamefont {N.}~\bibnamefont {Lorente}},\
  }\bibfield  {title} {\bibinfo {title} {Calculations of in-gap states of
  ferromagnetic spin chains on $s$-wave wide-band superconductors},\ }\href
  {https://doi.org/10.1103/PhysRevB.104.245415} {\bibfield  {journal} {\bibinfo
   {journal} {Phys. Rev. B}\ }\textbf {\bibinfo {volume} {104}},\ \bibinfo
  {pages} {245415} (\bibinfo {year} {2021}{\natexlab{b}})}\BibitemShut
  {NoStop}%
\bibitem [{\citenamefont {Jyoti}\ \emph {et~al.}(2024)\citenamefont {Jyoti},
  \citenamefont {Choi},\ and\ \citenamefont {Lorente}}]{Jyoti2024}%
  \BibitemOpen
  \bibfield  {author} {\bibinfo {author} {\bibfnamefont {D.}~\bibnamefont
  {Jyoti}}, \bibinfo {author} {\bibfnamefont {D.-J.}\ \bibnamefont {Choi}},\
  and\ \bibinfo {author} {\bibfnamefont {N.}~\bibnamefont {Lorente}},\
  }\bibfield  {title} {\bibinfo {title} {In-gap states induced by magnetic
  impurities on wide-band $s$-wave superconductors: Self-consistent
  calculations},\ }\href {https://doi.org/10.1103/PhysRevB.110.205404}
  {\bibfield  {journal} {\bibinfo  {journal} {Phys. Rev. B}\ }\textbf {\bibinfo
  {volume} {110}},\ \bibinfo {pages} {205404} (\bibinfo {year}
  {2024})}\BibitemShut {NoStop}%
\bibitem [{\citenamefont {Sticlet}\ \emph {et~al.}(2012)\citenamefont
  {Sticlet}, \citenamefont {Bena},\ and\ \citenamefont {Simon}}]{Sticlet}%
  \BibitemOpen
  \bibfield  {author} {\bibinfo {author} {\bibfnamefont {D.}~\bibnamefont
  {Sticlet}}, \bibinfo {author} {\bibfnamefont {C.}~\bibnamefont {Bena}},\ and\
  \bibinfo {author} {\bibfnamefont {P.}~\bibnamefont {Simon}},\ }\bibfield
  {title} {\bibinfo {title} {Spin and {Majorana} {Polarization} in
  {Topological} {Superconducting} {Wires}},\ }\bibfield  {journal} {\bibinfo
  {journal} {Physical Review Letters}\ }\textbf {\bibinfo {volume} {108}},\
  \href {https://doi.org/10.1103/PhysRevLett.108.096802}
  {10.1103/PhysRevLett.108.096802} (\bibinfo {year} {2012})\BibitemShut
  {NoStop}%
\bibitem [{\citenamefont {Bj\"ornson}\ and\ \citenamefont
  {Black-Schaffer}(2018)}]{Bjornson_2018}%
  \BibitemOpen
  \bibfield  {author} {\bibinfo {author} {\bibfnamefont {K.}~\bibnamefont
  {Bj\"ornson}}\ and\ \bibinfo {author} {\bibfnamefont {A.~M.}\ \bibnamefont
  {Black-Schaffer}},\ }\bibfield  {title} {\bibinfo {title} {Probing chiral
  edge states in topological superconductors through spin-polarized local
  density of state measurements},\ }\href
  {https://doi.org/10.1103/PhysRevB.97.140504} {\bibfield  {journal} {\bibinfo
  {journal} {Phys. Rev. B}\ }\textbf {\bibinfo {volume} {97}},\ \bibinfo
  {pages} {140504} (\bibinfo {year} {2018})}\BibitemShut {NoStop}%
\bibitem [{\citenamefont {Mashkoori}\ \emph {et~al.}(2020)\citenamefont
  {Mashkoori}, \citenamefont {Pradhan}, \citenamefont {Bj\"ornson},
  \citenamefont {Fransson},\ and\ \citenamefont {Black-Schaffer}}]{Mahdi2020}%
  \BibitemOpen
  \bibfield  {author} {\bibinfo {author} {\bibfnamefont {M.}~\bibnamefont
  {Mashkoori}}, \bibinfo {author} {\bibfnamefont {S.}~\bibnamefont {Pradhan}},
  \bibinfo {author} {\bibfnamefont {K.}~\bibnamefont {Bj\"ornson}}, \bibinfo
  {author} {\bibfnamefont {J.}~\bibnamefont {Fransson}},\ and\ \bibinfo
  {author} {\bibfnamefont {A.~M.}\ \bibnamefont {Black-Schaffer}},\ }\bibfield
  {title} {\bibinfo {title} {Identification of topological superconductivity in
  magnetic impurity systems using bulk spin polarization},\ }\href
  {https://doi.org/10.1103/PhysRevB.102.104501} {\bibfield  {journal} {\bibinfo
   {journal} {Phys. Rev. B}\ }\textbf {\bibinfo {volume} {102}},\ \bibinfo
  {pages} {104501} (\bibinfo {year} {2020})}\BibitemShut {NoStop}%
\bibitem [{\citenamefont {Moore}\ \emph {et~al.}(2018)\citenamefont {Moore},
  \citenamefont {Stanescu},\ and\ \citenamefont {Tewari}}]{Moore}%
  \BibitemOpen
  \bibfield  {author} {\bibinfo {author} {\bibfnamefont {C.}~\bibnamefont
  {Moore}}, \bibinfo {author} {\bibfnamefont {T.~D.}\ \bibnamefont
  {Stanescu}},\ and\ \bibinfo {author} {\bibfnamefont {S.}~\bibnamefont
  {Tewari}},\ }\bibfield  {title} {\bibinfo {title} {Two-terminal charge
  tunneling: Disentangling majorana zero modes from partially separated andreev
  bound states in semiconductor-superconductor heterostructures},\ }\href
  {https://doi.org/10.1103/PhysRevB.97.165302} {\bibfield  {journal} {\bibinfo
  {journal} {Phys. Rev. B}\ }\textbf {\bibinfo {volume} {97}},\ \bibinfo
  {pages} {165302} (\bibinfo {year} {2018})}\BibitemShut {NoStop}%
\bibitem [{\citenamefont {Herrera}\ \emph {et~al.}(2015)\citenamefont
  {Herrera}, \citenamefont {Guillam\'on}, \citenamefont {Galvis}, \citenamefont
  {Correa}, \citenamefont {Fente}, \citenamefont {Luccas}, \citenamefont
  {Mompean}, \citenamefont {Garcia-Hernandez}, \citenamefont {Vieira},
  \citenamefont {Brison},\ and\ \citenamefont {Suderow}}]{Herrera}%
  \BibitemOpen
  \bibfield  {author} {\bibinfo {author} {\bibfnamefont {E.}~\bibnamefont
  {Herrera}}, \bibinfo {author} {\bibfnamefont {I.}~\bibnamefont
  {Guillam\'on}}, \bibinfo {author} {\bibfnamefont {J.~A.}\ \bibnamefont
  {Galvis}}, \bibinfo {author} {\bibfnamefont {A.}~\bibnamefont {Correa}},
  \bibinfo {author} {\bibfnamefont {A.}~\bibnamefont {Fente}}, \bibinfo
  {author} {\bibfnamefont {R.~F.}\ \bibnamefont {Luccas}}, \bibinfo {author}
  {\bibfnamefont {F.~J.}\ \bibnamefont {Mompean}}, \bibinfo {author}
  {\bibfnamefont {M.}~\bibnamefont {Garcia-Hernandez}}, \bibinfo {author}
  {\bibfnamefont {S.}~\bibnamefont {Vieira}}, \bibinfo {author} {\bibfnamefont
  {J.~P.}\ \bibnamefont {Brison}},\ and\ \bibinfo {author} {\bibfnamefont
  {H.}~\bibnamefont {Suderow}},\ }\bibfield  {title} {\bibinfo {title}
  {Magnetic field dependence of the density of states in the multiband
  superconductor \protect{$\beta$-Bi$_2$Pd}},\ }\href
  {https://doi.org/10.1103/PhysRevB.92.054507} {\bibfield  {journal} {\bibinfo
  {journal} {Physical Review B}\ }\textbf {\bibinfo {volume} {92}},\ \bibinfo
  {pages} {054507} (\bibinfo {year} {2015})}\BibitemShut {NoStop}%
\bibitem [{\citenamefont {Schneider}\ \emph
  {et~al.}(2021{\natexlab{b}})\citenamefont {Schneider}, \citenamefont {Beck},
  \citenamefont {Neuhaus-Steinmetz}, \citenamefont {Posske}, \citenamefont
  {Wiebe},\ and\ \citenamefont {Wiesendanger}}]{Schneider2}%
  \BibitemOpen
  \bibfield  {author} {\bibinfo {author} {\bibfnamefont {L.}~\bibnamefont
  {Schneider}}, \bibinfo {author} {\bibfnamefont {P.}~\bibnamefont {Beck}},
  \bibinfo {author} {\bibfnamefont {J.}~\bibnamefont {Neuhaus-Steinmetz}},
  \bibinfo {author} {\bibfnamefont {T.}~\bibnamefont {Posske}}, \bibinfo
  {author} {\bibfnamefont {J.}~\bibnamefont {Wiebe}},\ and\ \bibinfo {author}
  {\bibfnamefont {R.}~\bibnamefont {Wiesendanger}},\ }\bibfield  {title}
  {\bibinfo {title} {Controlled length-dependent interaction of {Majorana}
  modes in {Yu}-{Shiba}-{Rusinov} chains},\ }\href
  {http://arxiv.org/abs/2104.11503} {\bibfield  {journal} {\bibinfo  {journal}
  {arXiv:2104.11503 [cond-mat]}\ } (\bibinfo {year} {2021}{\natexlab{b}})},\
  \bibinfo {note} {arXiv: 2104.11503}\BibitemShut {NoStop}%
\bibitem [{\citenamefont {Flatt\'e}\ and\ \citenamefont
  {Byers}(1997)}]{Flatte_1997}%
  \BibitemOpen
  \bibfield  {author} {\bibinfo {author} {\bibfnamefont {M.~E.}\ \bibnamefont
  {Flatt\'e}}\ and\ \bibinfo {author} {\bibfnamefont {J.~M.}\ \bibnamefont
  {Byers}},\ }\bibfield  {title} {\bibinfo {title} {Local electronic structure
  of a single magnetic impurity in a superconductor},\ }\href
  {https://doi.org/10.1103/PhysRevLett.78.3761} {\bibfield  {journal} {\bibinfo
   {journal} {Phys. Rev. Lett.}\ }\textbf {\bibinfo {volume} {78}},\ \bibinfo
  {pages} {3761} (\bibinfo {year} {1997})}\BibitemShut {NoStop}%
\bibitem [{\citenamefont {Flatt\'e}(2000)}]{Flatte_2000}%
  \BibitemOpen
  \bibfield  {author} {\bibinfo {author} {\bibfnamefont {M.~E.}\ \bibnamefont
  {Flatt\'e}},\ }\bibfield  {title} {\bibinfo {title} {Quasiparticle resonant
  states as a probe of short-range electronic structure and andre\'ev
  coherence},\ }\href {https://doi.org/10.1103/PhysRevB.61.R14920} {\bibfield
  {journal} {\bibinfo  {journal} {Phys. Rev. B}\ }\textbf {\bibinfo {volume}
  {61}},\ \bibinfo {pages} {R14920} (\bibinfo {year} {2000})}\BibitemShut
  {NoStop}%
\bibitem [{\citenamefont {Pientka}\ \emph {et~al.}(2013)\citenamefont
  {Pientka}, \citenamefont {Glazman},\ and\ \citenamefont {von
  Oppen}}]{Pientka2013}%
  \BibitemOpen
  \bibfield  {author} {\bibinfo {author} {\bibfnamefont {F.}~\bibnamefont
  {Pientka}}, \bibinfo {author} {\bibfnamefont {L.~I.}\ \bibnamefont
  {Glazman}},\ and\ \bibinfo {author} {\bibfnamefont {F.}~\bibnamefont {von
  Oppen}},\ }\bibfield  {title} {\bibinfo {title} {Topological superconducting
  phase in helical {Shiba} chains},\ }\href
  {https://doi.org/10.1103/PhysRevB.88.155420} {\bibfield  {journal} {\bibinfo
  {journal} {Physical Review B}\ }\textbf {\bibinfo {volume} {88}},\ \bibinfo
  {pages} {155420} (\bibinfo {year} {2013})}\BibitemShut {NoStop}%
\bibitem [{\citenamefont {Jeon}\ \emph {et~al.}(2017)\citenamefont {Jeon},
  \citenamefont {Xie}, \citenamefont {Li}, \citenamefont {Wang}, \citenamefont
  {Bernevig},\ and\ \citenamefont {Yazdani}}]{Yazdani2017}%
  \BibitemOpen
  \bibfield  {author} {\bibinfo {author} {\bibfnamefont {S.}~\bibnamefont
  {Jeon}}, \bibinfo {author} {\bibfnamefont {Y.}~\bibnamefont {Xie}}, \bibinfo
  {author} {\bibfnamefont {J.}~\bibnamefont {Li}}, \bibinfo {author}
  {\bibfnamefont {Z.}~\bibnamefont {Wang}}, \bibinfo {author} {\bibfnamefont
  {B.~A.}\ \bibnamefont {Bernevig}},\ and\ \bibinfo {author} {\bibfnamefont
  {A.}~\bibnamefont {Yazdani}},\ }\bibfield  {title} {\bibinfo {title}
  {Distinguishing a {Majorana} zero mode using spin-resolved measurements},\
  }\href {https://doi.org/10.1126/science.aan3670} {\bibfield  {journal}
  {\bibinfo  {journal} {Science}\ }\textbf {\bibinfo {volume} {358}},\ \bibinfo
  {pages} {772} (\bibinfo {year} {2017})}\BibitemShut {NoStop}%
\bibitem [{\citenamefont {Palacio-Morales}\ \emph {et~al.}(2019)\citenamefont
  {Palacio-Morales}, \citenamefont {Mascot}, \citenamefont {Cocklin},
  \citenamefont {Kim}, \citenamefont {Rachel}, \citenamefont {Morr},\ and\
  \citenamefont {Wiesendanger}}]{Wiesendanger2019}%
  \BibitemOpen
  \bibfield  {author} {\bibinfo {author} {\bibfnamefont {A.}~\bibnamefont
  {Palacio-Morales}}, \bibinfo {author} {\bibfnamefont {E.}~\bibnamefont
  {Mascot}}, \bibinfo {author} {\bibfnamefont {S.}~\bibnamefont {Cocklin}},
  \bibinfo {author} {\bibfnamefont {H.}~\bibnamefont {Kim}}, \bibinfo {author}
  {\bibfnamefont {S.}~\bibnamefont {Rachel}}, \bibinfo {author} {\bibfnamefont
  {D.~K.}\ \bibnamefont {Morr}},\ and\ \bibinfo {author} {\bibfnamefont
  {R.}~\bibnamefont {Wiesendanger}},\ }\bibfield  {title} {\bibinfo {title}
  {Atomic-scale interface engineering of majorana edge modes in a 2d
  magnet-superconductor hybrid system},\ }\href
  {https://doi.org/10.1126/sciadv.aav6600} {\bibfield  {journal} {\bibinfo
  {journal} {Science Advances}\ }\textbf {\bibinfo {volume} {5}},\ \bibinfo
  {pages} {eaav6600} (\bibinfo {year} {2019})}\BibitemShut {NoStop}%
\bibitem [{\citenamefont {Peters}\ \emph {et~al.}(2020)\citenamefont {Peters},
  \citenamefont {Bogdanoff}, \citenamefont {Acero~González}, \citenamefont
  {Melischek}, \citenamefont {Simon}, \citenamefont {Reecht}, \citenamefont
  {Winkelmann}, \citenamefont {von Oppen},\ and\ \citenamefont
  {Franke}}]{peters_resonant_2020}%
  \BibitemOpen
  \bibfield  {author} {\bibinfo {author} {\bibfnamefont {O.}~\bibnamefont
  {Peters}}, \bibinfo {author} {\bibfnamefont {N.}~\bibnamefont {Bogdanoff}},
  \bibinfo {author} {\bibfnamefont {S.}~\bibnamefont {Acero~González}},
  \bibinfo {author} {\bibfnamefont {L.}~\bibnamefont {Melischek}}, \bibinfo
  {author} {\bibfnamefont {J.~R.}\ \bibnamefont {Simon}}, \bibinfo {author}
  {\bibfnamefont {G.}~\bibnamefont {Reecht}}, \bibinfo {author} {\bibfnamefont
  {C.~B.}\ \bibnamefont {Winkelmann}}, \bibinfo {author} {\bibfnamefont
  {F.}~\bibnamefont {von Oppen}},\ and\ \bibinfo {author} {\bibfnamefont
  {K.~J.}\ \bibnamefont {Franke}},\ }\bibfield  {title} {\bibinfo {title}
  {Resonant {Andreev} reflections probed by photon-assisted tunnelling at the
  atomic scale},\ }\href {https://doi.org/10.1038/s41567-020-0972-z} {\bibfield
   {journal} {\bibinfo  {journal} {Nature Physics}\ }\textbf {\bibinfo {volume}
  {16}},\ \bibinfo {pages} {1222} (\bibinfo {year} {2020})},\ \bibinfo {note}
  {number: 12 Publisher: Nature Publishing Group}\BibitemShut {NoStop}%
\bibitem [{\citenamefont {González}\ \emph {et~al.}(2020)\citenamefont
  {González}, \citenamefont {Melischek}, \citenamefont {Peters}, \citenamefont
  {Flensberg}, \citenamefont {Franke},\ and\ \citenamefont {von
  Oppen}}]{gonzalez_photon-assisted_2020}%
  \BibitemOpen
  \bibfield  {author} {\bibinfo {author} {\bibfnamefont {S.~A.}\ \bibnamefont
  {González}}, \bibinfo {author} {\bibfnamefont {L.}~\bibnamefont
  {Melischek}}, \bibinfo {author} {\bibfnamefont {O.}~\bibnamefont {Peters}},
  \bibinfo {author} {\bibfnamefont {K.}~\bibnamefont {Flensberg}}, \bibinfo
  {author} {\bibfnamefont {K.~J.}\ \bibnamefont {Franke}},\ and\ \bibinfo
  {author} {\bibfnamefont {F.}~\bibnamefont {von Oppen}},\ }\bibfield  {title}
  {\bibinfo {title} {Photon-assisted resonant {Andreev} reflections:
  {Yu}-{Shiba}-{Rusinov} and {Majorana} states},\ }\href
  {https://doi.org/10.1103/PhysRevB.102.045413} {\bibfield  {journal} {\bibinfo
   {journal} {Physical Review B}\ }\textbf {\bibinfo {volume} {102}},\ \bibinfo
  {pages} {045413} (\bibinfo {year} {2020})},\ \bibinfo {note} {publisher:
  American Physical Society}\BibitemShut {NoStop}%
\bibitem [{\citenamefont {Kot}\ \emph {et~al.}(2020)\citenamefont {Kot},
  \citenamefont {Drost}, \citenamefont {Uhl}, \citenamefont {Ankerhold},
  \citenamefont {Cuevas},\ and\ \citenamefont
  {Ast}}]{kot_microwave-assisted_2020}%
  \BibitemOpen
  \bibfield  {author} {\bibinfo {author} {\bibfnamefont {P.}~\bibnamefont
  {Kot}}, \bibinfo {author} {\bibfnamefont {R.}~\bibnamefont {Drost}}, \bibinfo
  {author} {\bibfnamefont {M.}~\bibnamefont {Uhl}}, \bibinfo {author}
  {\bibfnamefont {J.}~\bibnamefont {Ankerhold}}, \bibinfo {author}
  {\bibfnamefont {J.~C.}\ \bibnamefont {Cuevas}},\ and\ \bibinfo {author}
  {\bibfnamefont {C.~R.}\ \bibnamefont {Ast}},\ }\bibfield  {title} {\bibinfo
  {title} {Microwave-assisted tunneling and interference effects in
  superconducting junctions under fast driving signals},\ }\href
  {https://doi.org/10.1103/PhysRevB.101.134507} {\bibfield  {journal} {\bibinfo
   {journal} {Physical Review B}\ }\textbf {\bibinfo {volume} {101}},\ \bibinfo
  {pages} {134507} (\bibinfo {year} {2020})},\ \bibinfo {note} {publisher:
  American Physical Society}\BibitemShut {NoStop}%
\end{thebibliography}%

\end{document}